\documentclass[sn-mathphys-ay]{sn-jnl}%

\usepackage{graphicx}%
\usepackage{multirow}%
\usepackage{amsmath,amssymb,amsfonts}%
\usepackage{amsthm}%
\usepackage{mathrsfs}%
\usepackage[title]{appendix}%
\usepackage[table,dvipsnames]{xcolor}
\usepackage{textcomp}%
\usepackage{manyfoot}%
\usepackage{booktabs}
\usepackage{multirow}
\usepackage{multicol}
\usepackage{array}
\usepackage{algorithm}%
\usepackage{algorithmicx}%
\usepackage{algpseudocode}%
\usepackage{listings}%

\usepackage{circledsteps}
\newcommand*\cnum[3]{
	\Circled[outer color={#1}, fill color={#1}, inner color={#2}]{\textbf{#3}}
}

\newcommand{\RNum}[1]{\uppercase\expandafter{\romannumeral #1\relax}}

\definecolor{olive}{RGB}{103, 120, 33}

\definecolor{CeruleanRef}{RGB}{12,127,172}
\hypersetup{colorlinks=true}
\hypersetup{allcolors=CeruleanRef}

\newcommand{\nsamples}{{M^\prime}}

\usepackage{subfig}
\usepackage[font=footnotesize,aboveskip=1pt,labelfont=bf,labelsep=period,justification=justified,singlelinecheck=off]{caption}

\theoremstyle{thmstyleone}%

\theoremstyle{thmstyletwo}%

\theoremstyle{thmstylethree}%

\begin{document}

\title[Article Title]{Calibration of stochastic, agent-based neuron growth models with Approximate Bayesian Computation}

\author*[1,2]{\fnm{Tobias} \sur{Duswald}}\email{\textcolor{CeruleanRef}{tobias.duswald@tum.de}}

\author[3]{\fnm{Lukas} \sur{Breitwieser}}

\author[4]{\fnm{Thomas} \sur{Thorne}}

\author[2]{\fnm{Barbara} \sur{Wohlmuth}}
\equalcont{These authors share senior authorship.}

\author[4]{\fnm{Roman} \sur{Bauer}}
\equalcont{These authors share senior authorship.}

\affil*[1]{\orgname{CERN}, \orgaddress{\country{Switzerland}}}
\affil[2]{\orgdiv{School of Computation, Information, and Technology},\orgname{Technical University of Munich}, \orgaddress{\country{Germany}}}
\affil[3]{\orgname{ETH Zurich}, \orgaddress{\country{Switzerland}}}
\affil[4]{\orgname{University of Surrey}, \orgaddress{\country{United Kingdom}}}

\abstract{
	Understanding how genetically encoded rules drive and guide complex neuronal growth processes is essential to comprehending the brain's architecture, and agent-based models (ABMs) offer a powerful simulation approach to further develop this understanding.
	However, accurately calibrating these models remains a challenge.
	Here, we present a novel application of Approximate Bayesian Computation (ABC) to address this issue.
	ABMs are based on parametrized stochastic rules that describe the time evolution of small components -- the so-called agents -- discretizing the system, leading to stochastic simulations that require appropriate treatment.
	Mathematically, the calibration defines a stochastic inverse problem.
	We propose to address it in a Bayesian setting using ABC.
	We facilitate the repeated comparison between data and simulations by quantifying the morphological information of single neurons with so-called \textit{morphometrics} and resort to \textit{statistical distances} to measure discrepancies between populations thereof.
	We conduct experiments on synthetic as well as experimental data.
	We find that ABC utilizing \textit{Sequential Monte Carlo} sampling and the \textit{Wasserstein distance} finds accurate posterior parameter distributions for representative ABMs.
	We further demonstrate that these ABMs capture specific features of pyramidal cells of the hippocampus (CA1).
	Overall, this work establishes a robust framework for calibrating agent-based neuronal growth models and opens the door for future investigations using Bayesian techniques for model building, verification, and adequacy assessment.
}

\keywords{Approximate Bayesian Computation, Neuronal Growth, Agent-Based Models, Calibration}

\pacs[MSC Classification]{62F15, 62F25, 92-04, 92-08, 92-10, 92B05, 92C10, 92C20, 92C42}

\maketitle

\section{Introduction}\label{sec1}

The brain is the human's most complex organ and comprises roughly 86 billion neurons~\citep{Azevedo2009, Herculano2009}, each of which is connected to hundreds or thousands of others via synapses. Brain regions accommodate different neuron types, providing specialized function for specific tasks. For example, the primate neocortex is made up of \textit{pyramidal cells} (70\%) as well as calretinin, calbindin, and parvalbumin-expressing \textit{interneurons} (30\%)~\citep{DeFelipe1992, Markram2004, Elston2011, DeFelipe1997, TorresGomez2020}. Researchers differentiate neurons by their location in the brain and their \textit{morphology}, i.e., their shape and form, but even neurons originating from the same species and brain region may show significant morphological differences~\citep{Deitcher2017}. Theoretical considerations, e.g., regarding the information capacity of the genome, have led various scholars to conclude that the brain's wiring and the neurons' morphologies likely emerge from simple developmental rules~\citep{Linsker1986,Hassan2015,Zador2019}. Mechanistic, agent-based neuron growth models hold the potential to investigate this hypothesis and unlock a deeper understanding of how neurons grow and build their elaborate networks.

A rich set of mathematical models has been established to capture the diverse properties of neurons, contributing to a more comprehensive understanding of these complex cells. Early research efforts focused on understanding responses to external electrical stimuli, leading to the development of influential models such as the Hodgkin-Huxley model~\citep{Hodgkin1952}. Simple, rate-based models capture the signal processing capabilities of neurons embedded in networks and became the workhorse of modern artificial intelligence applications, driving much of success in processing images and text~\citep{LeCun2015}. Conceptually different methods have been developed to recreate structures in line with the characteristic neuronal morphologies. Notable examples use L-systems~\citep{Lindenmayer1968,Hamilton1993,Ascoli2001}, statistically sample components from data~\citep{Nielsen2008}, or derive the structure from optimal wiring principles~\citep{Cuntz2010}. While these approaches successfully model the morphology, they provide limited insights into the fundamental processes driving the growth~\citep{Zubler2009}.

In this work, we focus on mechanistic, \textit{agent-based models} (ABM) simulating neuronal growth. In contrast to previously-mentioned approaches, such models are based on \textit{first principles}, act on \textit{local information}, and simulate growth in a \textit{biologically realistic manner}. The simulation begins with a simple initial configuration of a single neuronal soma. Afterward, the dendrites and the axons form and extend from the soma to shape the neuron. The models are based on discrete compartments, so-called agents. These agents act independently based on internal state variables and external, local information. Stochastic rules define their behavior, e.g., the rules may encode stochastic branching processes or random walk models. A comprehensive description of the mathematics governing these mechanistic ABMs may be found in the work of \citet{Zubler2009}. Ultimately, these models yield artificial neurons, which can be compared to real neurons. This comparison is non-trivial and carried out by reducing the neurons to their \textit{morphometrics}~\citep{Nielsen2014}, i.e., a set of \textit{quantities of interest} that capture the structural information of the neuron.

Early ABMs and related modelling studies for neuronal growth explored and explained various developmental aspects such as cell proliferation, polarization, and migrations~\citep{Ryder1999, Shinbrot2006, Samuels1996, Cai2006} as well as growth cone steering and neurite extension~\citep{Krottje2007, Goodhill2004, Kiddie2005}. Later works composed increasingly extensive models, e.g., \citet{Bauer2014a,Bauer2014b} presented models probing how connectivity arises in the neocortex and how first principles can lead to \textit{winner-takes-it-all} circuits. Using similar techniques, \citet{KassraianFrad2020} presented an ABM explaining axonal branching. More recently, \citet{Shree2022} observed growing neurons \textit{in-vivo} over time and derived a detailed ABM for sensory neurons.

While different growth models have been proposed, inferring their (latent) parameters presents a significant challenge because the models are strongly stochastic and data is usually limited. Mathematically, we face a stochastic inverse problem: given some data $y_{\mathit{obs}}$, find the distribution of the model parameters $p(\Theta | y_{\mathit{obs}})$ which best explain the data. Simple optimization-based procedures for identifying parameters may give false confidence in their values as they typically do not account for uncertainties. However, Bayesian methods have shown significant advances over the past three decades~\citep{Martin2020,Martin2024}, such that solving the inverse problem for neuronal structures in the presence of uncertainties is within reach.

In this work, we advocate embedding the growth models into a Bayesian framework to better understand the problem's characteristics, fusing ideas from theoretical neuroscience, computer science, and statistics to address the stochastic inverse problem. We describe an abstract concept termed \textit{resource-driven neuron growth model} motivated through experimental findings highlighting the effect of the neuron's transport system on its morphology and consider two simple representatives thereof. We propose using \textit{Approximate Bayesian Computation} (ABC)~\citep{Tavare1997, Pritchard1999, Beaumont2002, Marjoram2003, Csillery2010, Beaumont2010, Sisson2007, Sisson2019} combined with a selected set of morphometrics and statistical distances~\citep{Bernton2019,Kimia2020,Jiang2018,Fujisawa2021} to find approximations to the \textit{posterior distribution} $p(\Theta | y_{\mathit{obs}})$. To this end, we employ \textit{Del Moral's SMCABC algorithm}~\citep{DelMoral2012} based on \textit{sequential Monte Carlo} (SMC) sampling~\citep{DelMoral2006} with modifications to the kernel and the distance metric as proposed by~\citet{Bernton2019}. These modifications allow us to bypass the definition of \textit{summary statistics} and measure the discrepancy between data and simulation directly with the \textit{Wasserstein distance} (or similar). The algorithm is inherently parallel and, thus, scalable on modern computing resources. We leverage the MPI-parallel implementation of ABCpy~\citep{Dutta2021} and embed computational models implemented with the highly efficient BioDynaMo~\citep{Breitwieser2021, Breitwieser2023} framework into the inner loop of the algorithm resulting in a high-throughput implementation. The data from the study is retrieved from neuromoropho.org~\citep{NeuroMorpho}, the most extensive database for neuronal morphologies. We demonstrate that the method can find adequate posterior distributions through computational experiments on synthetic data, i.e., data generated via the models, and subsequently show that the models can simulate pyramidal cell morphologies in agreement with experimental data collected by \citet{DeFelipePyramidal2019}. We share our implementation (see~\nameref{suppl}), and future research may leverage the framework to calibrate different models efficiently.

The article is structured as follows. We begin with a technical overview of the project, briefly explaining which components are relevant and how they connect. Afterward, we explain them in detail, i.e., we discuss neuron growth models, data sources, sensitivity analysis, and ABC algorithms. We then show numerical experiments that investigate the models' stochastic components and sensitivities before approaching the inverse problem with synthetic and experimental data. On synthetic data, we explore how the choice of morphometrics, statistical distances, and sample size affect the algorithm. We conclude the manuscript by critically reflecting on the results and embedding the findings in a broader context.

\section{Materials and methods}

Our principal goal is to determine the parameters and uncertainties of mechanistic neuron growth models for given data. We choose to address the problem in a Bayesian setting. Given a stochastic, computational model parameterized through parameters $\Theta \in \mathbb{R}^N$, we strive to find the \textit{posterior distribution} $p(\Theta|y_{\mathit{obs}})$, which is the probability distribution describing the parameters $\Theta$ after observing data $y_{\mathit{obs}}$. The posterior encapsulates all available information on the parameters, including the most probable values and associated uncertainties. Formally, the solution to this problem is given by Bayes' theorem
\begin{equation}\label{eq:bayes_theorem}
	p(\Theta|y_{\mathit{obs}}) = \frac{p(y_{\mathit{obs}} | \Theta) p(\Theta)}{p(y_{\mathit{obs}})} \, ,
\end{equation}
which defines the posterior in terms of
the following three components: the probability distribution of the
observed data $p(y_{\mathit{obs}})$, the \textit{prior distribution} of the
parameters $p(\Theta)$ containing all available knowledge about the parameters
prior to the calibration, and the \textit{likelihood function} $p(y_{\mathit{obs}} |
	\Theta)$ which describes how likely observations $y_{\mathit{obs}}$ are under given
parameters $\Theta$.
In practice, i.e., in all but the most straightforward cases, finding a closed-form solution to \eqref{eq:bayes_theorem} is impossible.
Hence, we numerically approximate the posterior distribution with algorithms whose details are presented and discussed in later sections.

\begin{figure}
	\centering
	\includegraphics[width=1.0\textwidth]{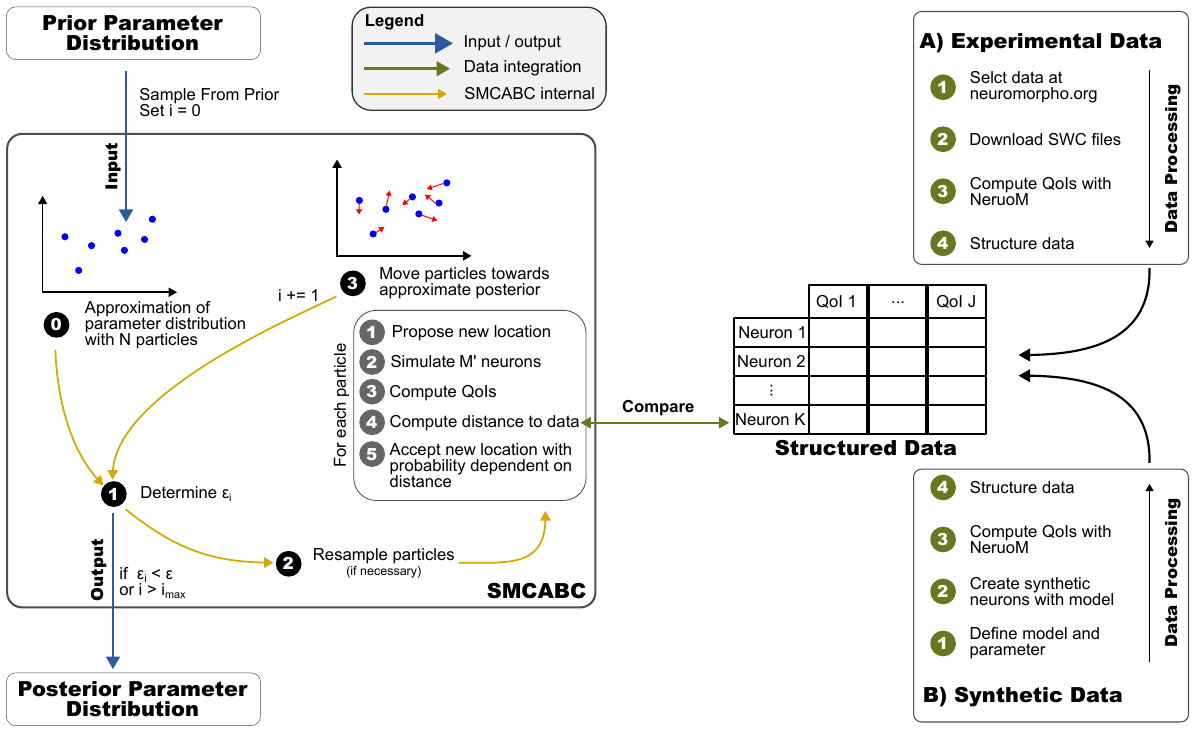}
	\caption{Conceptual overview of the calibration methodology.
		We first define the quantities of interest (QoI $1,\dots,J$) for the neuronal growth simulation.
		The data is then obtained from either of two data pipes: A) experimental data or B) synthetic data (\cnum{olive}{white}{1} to \cnum{olive}{white}{4}).
		After processing and structuring the data ($J$ QoIs for $K$ neurons), we provide an initial guess of the model parameters -- the prior parameter distribution -- to the SMCABC algorithm~\citep{DelMoral2012,Bernton2019}.
		The algorithm~\citep{DelMoral2012} describes the parameter distribution with a set of $N$ particles (\cnum{black}{white}{0}) and enters into a loop (\cnum{black}{white}{1} to \cnum{black}{white}{3}) iteratively moving particles closer to the posterior.
		Adaptively lowering the tolerance level $\epsilon_i$~(\cnum{black}{white}{1}) ensures efficient positional updates of the particles in~(\cnum{black}{white}{3})~\citep{DelMoral2012}.
		For such updates, the algorithm executes the steps \cnum{gray}{white}{1} to \cnum{gray}{white}{5} simulating $\nsamples$ neurons under the model and computing the statistical distance to the data.
	}
	\label{fig:project-overview}
\end{figure}

As we face a stochastic inverse problem, our study consists of four major components: mechanistic neuron growth models~(Section~\ref{methods:models}), data sources and processing~(Section~\ref{methods:data}), a numerical method solving~\eqref{eq:bayes_theorem}~(Section~\ref{methods:numerics}), and efficient software implementations and interfaces~(Section~\ref{fig:software-implementation}). In the following, we detail the different components and their links. Figure~\ref{fig:project-overview} gives an overview and shows how the components interact.

\subsection{Mechanistic Neuron Growth Models}\label{methods:models}

\begin{figure}
	\centering
	\includegraphics[width=\textwidth]{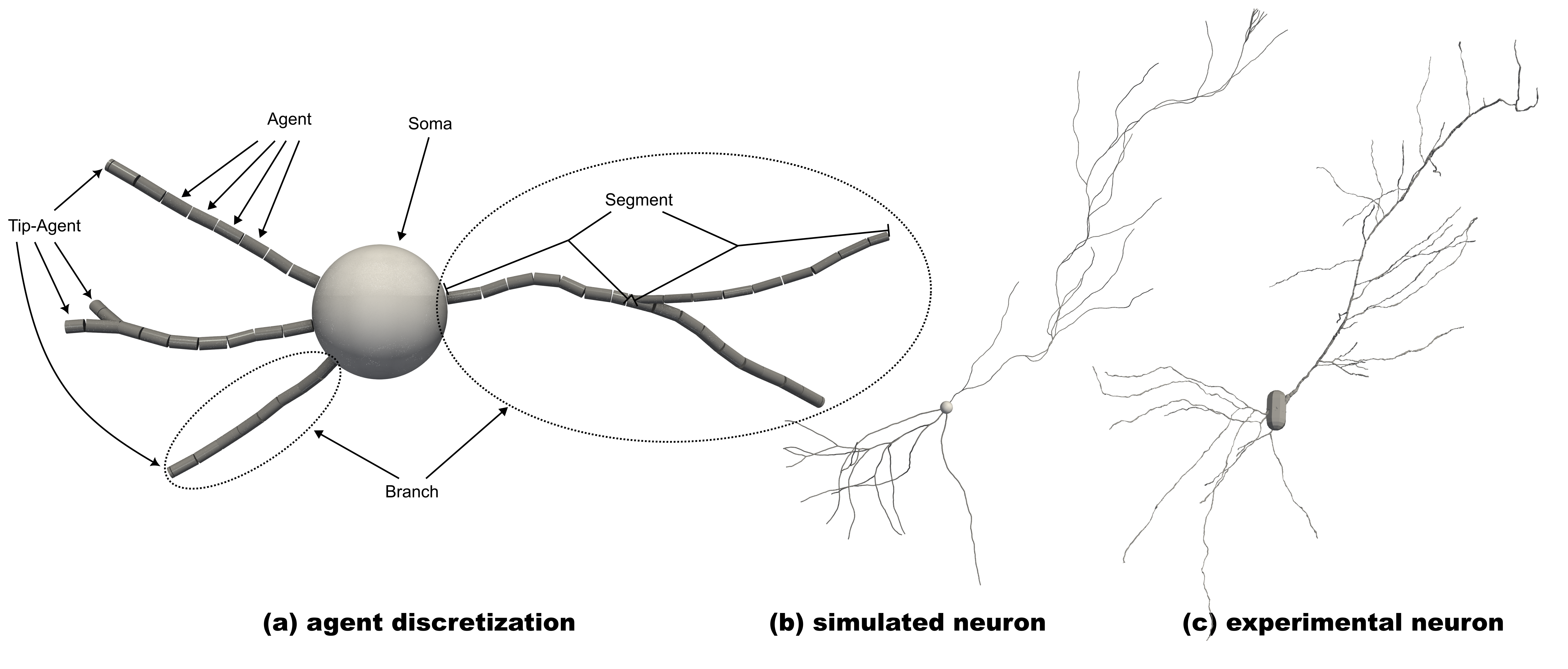}
	\caption{
		3D Mechanistic ABM for neuronal growth.
		Panel~(a) shows the agent-based discretization and the early stage of a neuron simulation (pyramidal cell).
		The center is the spherical soma (cell body).
		The dendrites are discretized with small, cylindrical agents.
		Typically, the tips drive the growth~\citep{Shree2022}; hence, we differentiate between the general and tip agents.
		The agents define a tree-like structure with different \textit{segments} and \textit{branches}.
		(b)~Pyramidal cell at the end of the simulation.
		(c)~Experimental pyramidal cell in the mouse hippocampus observed by~\citet{DeFelipePyramidal2019} (\href{https://neuromorpho.org/neuron_info.jsp?neuron_name=24-mouse-c57-hcca1-id6-sec53-cel3-soma-corr1-6z}{ NMO\_147071}).
	}
	\label{fig:neuron-abm}
\end{figure}

This work is concerned with 3D agent-based, mechanistic neuron growth models. Figure~\ref{fig:neuron-abm} displays an example of such a simulation. Panel~(a) illustrates the ABM-discretization and the simulated neuron in an early stage of the simulation, panel~(b) depicts the final simulated neuron, and (c) shows an experimentally observed pyramidal cell in the mouse hippocampus~\citep{DeFelipePyramidal2019}. The initial conditions of the simulation are not displayed but may be obtained by stripping all agents from (a) not directly attached to the soma. We will first introduce some biological background, continue presenting the ABM discretization, and then explain the high-level abstraction of a resource-driven neuron growth model. Lastly, we consider two simple model realizations corresponding to the upper and lower parts of panel (b) relevant to the subsequent numerical experiments.

\subsubsection{Biological background}

Despite a good understanding of neurons properties and their networks, the processes by which neurons develop their morphologies are only partially comprehended. A neuron's journey begins during \textit{neurogenesis} - the process in which new neurons form from neural stem or progenitor cells. Afterward, the neuron begins to extend dendrites and axons from the soma. The tips of dendrites and axons display a growth cone, a highly polarized, dynamic structure that guides the growth. External substances serve as attracting and repulsive guidance cues. Examples include netrins, semaphorins, ephrins, and the brain-derived neurotrophic factor. Neurons integrate into complex networks and, depending on the network's need, they may begin extending or retracting parts of their dendritic or axonal outgrowth.

Ultimately, growth requires resources; consequently, the morphology strongly depends on how resources get distributed within the neuron. The neuron's transport system is primarily built from three superfamilies of so-called \textit{molecular motor proteins}: kinesins, cytoplasmic dynein, and myosins. These proteins convert energy stored in adenosine triphosphate into mechanical energy, which they use to move along the cell's cytoskeletal tracks carrying cargo. \textit{Tubulin} is a cargo often considered in this context since it is a building block for the cytoskeleton and, thus, required for growth. It has been the driving factor for a set of mathematical models in the past~\citep{VanOoyen2001,McLean2004,Graham2006,Hjorth2014}. The influence of the motor proteins has been extensively reviewed by \citet{Hirokawa2010}, concluding that they play a significant role in the intercellular transport, control of neuronal function, morphogenesis, and survival. They also point out that expression levels differ from neuron to neuron~\citep{Kamal2000}, hinting that transport systems differ between neurons. Experimental evidence appears in multiple studies. \citet{Satoh2008} showed that gene mutations affecting the motor proteins lead to significant morphological changes in the dendritic arbor in \textit{drosophila melanogaster}, e.g., an overall shorter arbor and different branching pattern. \citet{Zheng2008} independently observed identical effects. Moreover, \citet{Hirokawa2009} demonstrated that suppressing \text{Kif2a}, part of the kinesin superfamily, caused abnormal axon branching, leading to significantly more branching points in the axon, see~\cite[Figure~5c]{Hirokawa2009}. \citet{Ryu2006} observed that \textit{myosin IIB} influences the morphology of dendritic spines, and \citet{Lyons2009} found that suppression of \textit{Kif1b} hinders the normal outgrowth of axons in zebrafish. We conclude that differences in the generation and distribution of resources significantly impact the morphology of different neurons. This should be reflected in the mathematical models and the resources become key-attributes for the agents.

\subsubsection{Agent-based neuron discretization}

We consider growth that starts from a few neurites attached to the soma, where the latter is stationary, i.e., it does not change throughout the simulation. The neurites are spatially discretized into cylindrical agents representing small dendritic tree compartments; recall Figure~\ref{fig:neuron-abm}~(a). The agents reside in a tree-like data structure, i.e., each agent has one mother and either zero, one, or two daughters. If an agent has no daughter, we refer to it as a tip agent. These are particularly important because the tips primarily drive the growth~\citep{Shree2022}. For all models, a cylindrical agent is characterized by its position, i.e., the start and end point of the cylinder, as well as their diameter. The orientation and length are implicitly contained in the start and end points.

The total simulation time $T$ is discretized into small time steps $\Delta t$ of equal duration. During each time step, the agents independently execute their stochastic rules governing their behavior. These rules may depend on local information, such as substance concentrations or gradients. Moreover, the rules may further depend on whether the agent has daughters; in other words, the rules may differ between regular and tip agents. It is desireable to parametrize the rules such that different choices for $\Delta t$ yield statistically identical (or at least similar) results. We further note that the stochastic processes encoded in the rules may restrict $\Delta t$ to a specific range.

This modeling approach is generally considered to be biologically feasible because all agents act on locally available information rather than globally optimizing specific properties. Algorithm~\ref{alg:BDM} in \nameref{S1_Algorithms} gives an overview of the simulation logic of ABMs implemented with BioDynaMo~\citep{Breitwieser2021}. By implementing the models with BioDynaMo, we implicitly include the neurons' mechanistic properties into the model. In this work, these properties play a minor role and we refer to~\citet{Zubler2009} for additional details.

\subsubsection{Resource-driven neuron growth model}

We introduce a high-level model description termed the \textit{resource-driven neuron growth model}. It is rooted in the realization that the molecular motors and further transport-related quantities differ in their expression levels between neuron types and the fact that they influence the morphology. Various models presented in the literature, e.g.,~\citep{Shree2022,KassraianFrad2020}, fit within the descriptions; here, we attempt to phrase it in a general parametric way.

The model needs to account for (1) the migration of the tips (elongation and retraction), (2) external guidance cues, (3) resource availability and transport, and (4) rules for branching and bifurcating. Different external guidance cues are described with scalar fields $\phi_i(x,t)$, and different resources of the agent-$j$ are denoted as $r_i^{(j)}$.

\citet{Shree2022} observed that tips either elongate, stall, or retract.
They suggest a model defining the stochastic transitions between the three states.
Additionally, they assume that contact of tips with other neurites causes them to retract.
When the tip retracts, it simply follows the path; when elongating, the tips execute a persistent, biased, random walk (biased through the gradient of external guidance cues; see~\citep{Codling2008,Hannezo2017} for details on such walks).
We add that, in general, the state transitions may depend on resource availability.
Furthermore, the state may influence the resource availability, e.g., it seems natural to assume that elongation reduces and retraction frees resources.

The time dependent resource distribution can be modeled in vastly different ways ranging from heuristic rules~\citep{KassraianFrad2020,VanOoyen2001} defining how the resources propagate when branching or bifurcating to intricate transport equations~\citep{Graham2006b,McLean2004,Hjorth2014,Qian2022}. Both approaches may capture the branches' competition for resources. Generally, branching is modelled as a (Poisson-like) stochastic process and the probability of branching per time step may depend on resource availability or external guidance cues. For example, the branching probability may increase with decreasing resource, leading to more, smaller branches towards the distal end.

These high-level requirements allow the construction of complex models with a high degree of incorporated biological information. Nonetheless, the key concern of this work is approximately solving the stochastic inverse problem, and we do not strive to create the biologically most detailed model. We therefore restrict ourselves two particularly simple representatives of such a model, which trace back to~\citep{Zubler2009,Breitwieser2021}. Both models use a single resource type, one guidance cue, and only consider the elongating and idle states. We describe the algorithm assuming some (simple) initial neuron structure is present, e.g., a structure as in Figure~\ref{fig:neuron-abm}~(a).

\paragraph{Example: Model 1.}
We first note that only tip-agents actively change; others remain untouched. Furthermore, only tip-agents whose resource satisfies $r > r_{\min}$ change; thus, if the resource of a tip-agent falls below the threshold $r_{\min}$, it becomes idle as well. In other cases (tip-agent with a sufficiently large resource), the agent elongates in the direction of a vector $\vec{d}$ which is composed out of (1)~a random component, (2)~the current orientation, and (3)~the direction of $\nabla \phi$ at the agent's position. Elongation means that the endpoint of the cylinder is shifted by $v \cdot \vec{d} / || \vec{d} ||_2$, where $v$ is the \textit{elongation speed parameter}. Since the elongation models a notion of stretching, the resource is decreased, i.e., $r(t_i + \Delta t) = r(t_i) - R$, where $R$ is the \textit{resource consumption parameter}. Each tip-agent can also branch with a constant probability $p_{bra}$. The two daughters created during branching inherit the resources of the mother.

A simulation of the branching process is depicted in Figure~\ref{fig:neuron-abm}~(b), where the structures below the soma are generated with Model~1. The symmetric resource distribution yields a balanced tree structure, e.g., two daughters of the same mother progress similarly. A pseudo-code representation of this description is given in \nameref{S1_Algorithms}, Algorithm~\ref{alg:Model1}.

\paragraph{Example: Model 2.}
Model 2 is similar to Model~1. In contrast to Model~1, all agents in Model~2 keep decreasing their resources until they reach or fall (slightly) below $r_{\min}$. Additionally, the branching rules differ, i.e., the neurite continues in a straight line and adds a new branch rather than symmetrically splitting into two. Sometimes this behavior is referred to \textit{branching}, while Model~1's symmetric splitting is referred to as \textit{bifurcating}. The asymmetry of the branching also reflects in how the resource is distributed; while the agent of the extended branch inherits the resource of the mother, the agent of newly-created branch initializes it to a fixed resource value of $r_0$, an additional parameter of Model~2. Hence, the resource distribution is asymmetric. We refer to~\citet{KassraianFrad2020} for a more involved asymmetric model.

While the differences between Model 1 and 2 are subtle, these differences may result in vastly different structures, e.g., Figure~\ref{fig:neuron-abm}~(b) shows the structures arising from Model~2 above the soma. The asymmetry of the resource distribution while branching leads to a more extended main branch with different shorter outgrowths. A pseudo-code representation of this description is given in \nameref{S1_Algorithms}, Algorithm~\ref{alg:Model2}.

\subsection{Data sources and processing}\label{methods:data}

In this study, we consider two types of datasets: synthetic and experimental data. The former serves as a test case for the algorithm and the choice of morphometrics; the latter is used to identify which models can describe the morphology of real neurons. The comparison of different neuronal structures is facilitated via the morphometrics.

\subsubsection{Morphometrics}

Morphometrics generally refers to the study of the size and shape of objects. We restrict ourselves to neurons and use the term morphometrics to describe features that quantify the morphology. The morphometrics attempt to answer the following question: How can we map a given neuron morphology onto a vector $x \in \mathbb{R}^n$ that adequately characterizes the neuron? This question is inherently challenging because neurons show very complicated shapes and forms. A short overview of popular morphometrics may, for instance, be found in the work of~\citet[Table~1.1]{Nielsen2014} or~\citet[Table~2]{Deitcher2017}. Popular features are total length, number of branches, mean branching length, and many more. Table~\ref{tab:overview-morphometrics} gives an overview of the simple morphometrics used in this work.

\begin{table}[]
	\centering
	\footnotesize
	\begin{tabular}{@{}ll@{}}
		\toprule
		ID             & Morphometric                         \\ \midrule
		$\mathcal{M}$1 & number of segments                   \\
		$\mathcal{M}$2 & mean segment length                  \\
		$\mathcal{M}$3 & standard deviation of segment length \\
		$\mathcal{M}$4 & total dendritic length               \\ \bottomrule
	\end{tabular}
	\caption{Overview of the subset of morphometrics used in this work.}
	\label{tab:overview-morphometrics}
\end{table}

If we abstractly consider the morphometrics (or a combination thereof) as a mapping $\mathcal{M}$ projecting from the space of neuronal morphologies $\mathcal{N}$ into $\mathbb{R}^n$, i.e., $\mathcal{M} : \mathcal{N} \rightarrow \mathbb{R}^n$, then the mapping does not possess an inverse. In other words, it is impossible to reconstruct the morphology from the morphometrics; however, similar neurons map to points that are close in $\mathbb{R}^n$. In general, the mapping consists of a combination of morphometrics. Describing the different morphometrics (e.g., branching length) as a map $\mathcal{M}_i : \mathcal{N} \rightarrow \mathbb{R}^{n_i}$, we define the morphometrics mapping as
\begin{equation}\label{eq:morphometrics-mapping}
	\mathcal{M} = \mathcal{M}_1 \times \dots \times \mathcal{M}_k
	: \mathcal{N} \rightarrow \mathbb{R}^n = \mathbb{R}^{n_1 + \dots + n_k} \, .
\end{equation}
While the mathematical nature of $\mathcal{M}$ may be complicated, the algorithmic implementation for computing $x = \mathcal{M}_i ( y \in \mathcal{N})$ is usually straightforward.
Following the jargon of predictive computational sciences, we will refer to $x$ as \textit{quantity of interest} (QoI) and use this term interchangeably with the morphometrics.
\subsubsection{Synthetic data}

We choose a stochastic computational model to generate a synthetic dataset by repeatedly executing it with different random seeds. $M$ model runs will result in $M$ distinctively different neuron samples. For each sample, we compute the morphometrics either with custom, unit-tested C++ code integrated into the model evaluation or offline with NeuroM~\citep{NeuroM}. This Python package allows the analysis of neuron morphologies saved in the SWC format and the extraction of their morphometrics. Finally, we structure the data in a spreadsheet-like data structure (see Figure~\ref{fig:project-overview}) to conveniently retrieve a given neuron's morphometrics.

\subsubsection{Experimental data}

\begin{table}[]
	\centering
	\footnotesize
	\begin{tabular}{@{}llllll@{}}
		\toprule
		dataset ID     & cardinality & neuron type & species & brain region      & reference                      \\ \midrule
		$\mathcal{D}$1 & 50          & pyramidal   & human   & hippocampus (CA1) & ~\citep{DeFelipePyramidal2019} \\
		$\mathcal{D}$2 & 50          & pyramidal   & mouse   & hippocampus (CA1) & ~\citep{DeFelipePyramidal2019} \\ \bottomrule
	\end{tabular}
	\caption{
		Overview of the experimental datasets retrieved from neuromorpho.org~\citep{NeuroMorpho}.
	}
	\label{tab:overview-data}
\end{table}

For the experimental data, we proceed akin to the synthetic data. We retrieve the data from neuromopho.org~\citep{NeuroMorpho}, an online database storing roughly 260 thousand digital reconstructions of neurons (as of October 2023). Typically, we select specific experiments and references or queries based on neuron type, species, and brain region. NeuroMorpho provides full access to the data; thus, we retrieve a set of neuronal morphologies in the SWC file format after selection. We use NeuroM~\citep{NeuroM} to verify that the morphologies are correct and that no errors in the file could harm the results. Subsequently, we extract the morphometrics and organize the data in a spreadsheet format. We consider two different datasets for our computational experiments. Table~\ref{tab:overview-data} gives an overview of them.

\subsection{Numerical methods}\label{methods:numerics}

The third cornerstone of this work are numerical methods fostering the understanding of the model's inherent stochasticity as well as those (approximately) solving the statistical inverse problem defined in~\eqref{eq:bayes_theorem}. We use sensitivity analysis algorithms to understand the models' behavior due to parameter variations and verifying which QoIs are informative for the parameter inference. The inference is conducted with Approximate Bayesian Computation.

\subsubsection{Sensitivity analysis}

Sensitivity analysis (SA) measures how strongly individual parameter influence the prediction of specific QoIs of a complex mathematical model~\citep{Saltelli2008}. In this work, we employ a global Sobol SA. We first define the (bounded) parameter space $\Omega$ for a given model. We use Saltelli's method~\citep{Saltelli2002, Saltelli2010} to draw $K$ samples from $\Omega$. Afterward, we use the model to generate a synthetic dataset of $\nsamples$ neurons for each sample drawn. We determine the QoIs and subsequently compute their expected value $\mathbb{E}[\mbox{QoI}]$. We use Sobol's method~\citep{Sobol2001} to estimate the sensitivity indices $S_1$ and $S_{tot}$ together with their 95\% confidence intervals. Here, $S_1$ is the index indicating how much of the variance in a QoI can be attributed to a given parameter (first-order sensitivity). The index $S_{tot}$ accumulates the first-order and higher-order indices to give an idea of the importance of a parameter; it is called the total-effect index. As the sensitivity measures variance, it may indicate which QoIs contain relevant information for parameter inference.

\subsubsection{Bayesian Computation}

Bayesian computation attempts to find numerical solutions and approximations to the Bayesian inverse problem~\eqref{eq:bayes_theorem}. Prominent candidates are Markov Chain Monte Carlo (MCMC) methods~\citep{Brooks2011} such as the Metropolis--Hastings algorithm~\citep{Metropolis1953,Hastings1970} or Gibbs sampling~\citep{Geman1984,Gelfand1990}. These methods are, however, limited to a small subset of real-world problems -- the ones with tractable likelihood $p(y_{\mathit{obs}} | \Theta)$. Nevertheless, many meaningful problems have an intractable likelihood, i.e., there may not be a closed form, or it may be too expensive to evaluate. This realization gave rise to a set of methods commonly referred to as \textit{likelihood-free methods}. Instead of evaluating the likelihood, the algorithms in this category operate under the assumption that simulating data under the model (or a surrogate thereof) facilitates an understanding of the likelihood. Representatives for these algorithms are Bayesian synthetic likelihood~\citep{Price2018}, specific versions of Variational Bayes~\citep{Beal2003,Jordan1999,Blei2017}, Integrated nested Laplace~\citep{Rue2009}, and, possibly the most popular one, Approximate Bayesian computation (ABC)~\citep{Tavare1997, Pritchard1999, Beaumont2002, Marjoram2003, Csillery2010, Beaumont2010, Sisson2007, Sisson2019}. In this work, we focus exclusively on ABC, which has proven to facilitate successful calibration in the context of ABMs in biological applications, e.g., \citep{Lambert2018,Wang2024}. For more information on the historical development of Bayesian computation, we refer to~\citet{Martin2020,Martin2024} and~\citet[Chapter~2]{Sisson2019}.

\subsubsection{Approximate Bayesian Computation}

ABC is based on the fundamentally simple idea of simulating data $y_{\mathit{sim}}$ under the model and comparing its output against the observed data $y_{\mathit{obs}}$. ABC algorithms must find $N$ simulations close to the data to obtain a Monte Carlo approximation to the posterior. Whether samples are accepted or not depends on a criterion involving a distance metric $d$ and a function $\eta: \mathbb{R}^n \rightarrow \mathbb{R}^m$ ($m \ll n$) called \textit{summary statistics}\footnote{The ABC literature often refers to QoIs that summarize the statistical characteristics of a simulation (e.g., $\mathcal{M}2$) as \textit{summary statistics}. To avoid any ambiguity in the terminology, we refer to values quantifying the simulated system's properties and statistics as \textit{QoIs} or \textit{morphometrics} and use the term \textit{summary statistics} exclusively for $\eta$. }. Simulations are considered close if
\begin{equation}\label{eq:abc_distance_criterion}
	d(\eta(y_{\mathit{obs}}), \eta(y_{\mathit{sim}})) < \epsilon \, ,
\end{equation}
where $y_{\mathit{obs}}$ and $y_{\mathit{sim}}$ denote datasets, i.e, they contain $M$ and $\nsamples$ $k$-dimensional random variables, respectively.
The function $\eta$ summarizes their statistics, allowing us to search for close points in a lower dimensional space, significantly speeding up the search.
For instance, we may calibrate the parameter $m$ of a Gaussian model $y \in \mathbb{R} \sim \mathcal{N}(m,\sigma)$ by choosing $\eta = \left(\sum_{i=1}^\nsamples y_i\right)/\nsamples$ since the mean adequately summarizes the statistics of the data.

Choosing appropriate or even sufficient summary statistics for arbitrary models remains one of the biggest challenges when employing ABC in practice. To overcome this limitation, \citet{Bernton2019} suggested using the~\textit{Wasserstein distance} to directly measure the discrepancy between simulated and observed data. Their approach generalizes the use of order statistics to arbitrary dimensions. The Wasserstein distance between two probability distributions measures how much work is necessary to turn one into the other. Hence, it is often called earth-movers distance and is computationally related to optimal transport problems. The distance is sometimes also called the \textit{Kantorovich–Rubinstein metric}. Other authors promoted similar ideas around the same time: \citet{Park2016} suggested using \textit{MMD}, \citet{Genevay2018} used \textit{Sinkhorn divergences}, and \citet{Jiang2018} employed the \textit{Kullback-Leibler divergence}. Later work considered the \textit{sliced-Wasserstein distance}~\citep{Kimia2020} and $\gamma$\textit{-divergence}~\citep{Fujisawa2021} targeting certain shortcoming of other distances.

The different statistical distances have similar effects in the ABC context but their interpretation differs; for instance, KL and $\gamma$ divergence measure the information loss when one distribution is used to approximate another. Moreover, their naming convention highlights a subtile mathematical difference; the Wasserstein distance is a \textit{metric} while KL and $\gamma$ represent a \textit{divergence}. The former is therefore non-negative, symmetric, and satisfies the triangle equation while the latter share the non-negativity but are non-symmetric and do not obey the triangle equality~(e.g.,~\cite[Chapter~1]{Amari2016}). Their different definitions yield qualitatively different behavior; e.g., considering two multivariate normal distributions $\mathcal{N}_1(\mu_1,\Sigma_1)$ and $\mathcal{N}_2(\mu_2,\Sigma_2)$, the Wasserstein distance scales linearly with $||\mu_1 - \mu_2||_2$~\citep{Dowson1982}, whereas straightforward calculation shows that the KL divergence scales quadratically with it.

Besides defining appropriate metrics for comparing simulated and observed data, designing algorithms that efficiently propose suitable samples has been a long-standing challenge in ABC. Over the past two decades, researchers derived many different ABC algorithms~\citep{Csillery2010, Sisson2019} and Sequential Monte Carlo (SMC) samplers~\citep{DelMoral2006, DelMoral2012} became a potent tool. SMC samplers represent the parameter distribution with $N$ \textit{particles} in the corresponding vector space. Instead of directly moving from the prior to the posterior, SMC algorithms propagate the particles through many intermediate probability distributions that change slowly from iteration to iteration, keeping the sampling efficient. Effectively, the sequence of distributions corresponds to a sequence of thresholds in the acceptance criterion \eqref{eq:abc_distance_criterion}, i.e., the sampler sequentially moves through distributions defined by the thresholds
\begin{equation}
	\epsilon_0 = \infty > \epsilon_1 > \dots > \epsilon_{k} = \epsilon \, .
\end{equation}
First, the particles are sampled from the prior ($\epsilon_0$).
By reducing $\epsilon_i$ from iteration to iteration, the posterior approximation through particles improves from iteration to iteration until eventually reaching the prescribed quality defined via $\epsilon_{k}$.
Del Moral's algorithm~\citep{DelMoral2006, DelMoral2012} bypasses the \textit{a priori} definition of the approximation levels $\epsilon_{i}$ by demanding that the effective sample size (ESS) of iteration $i+1$ is a certain fraction $\alpha \in (0,1)$ of the ESS of iteration $i$.
The threshold $\epsilon_{i+1}$ can be adaptively computed from the ESS, $\epsilon_{i}$, and $\alpha$ (see~\cite[Eq.~12]{DelMoral2012} for details).

To this end, we use Del Moral's SMCABC algorithm~\citep{DelMoral2006,DelMoral2012} with modifications proposed by \citet{Bernton2019}, i.e., we favor statistical distances over summary statistics and choose the arguably more efficient r-hit kernel ($r=2$)~\citep{Lee2012,Lee2014}. We choose to use $N=2^{10}$ particles and $\alpha = 0.6$.

To assess the quality of the resulting posterior distribution, we perform a \textit{predictive check}. This check involves drawing samples from the posterior, evaluating the model for each sample, and computing the QoIs. We then compare the simulated QoIs to the data; since the QoIs form a high dimensional space, we show the marginals of the distributions, i.e., the projection on one coordinate axis in the QoI space.

\begin{figure}
	\centering
	\includegraphics[width=\textwidth]{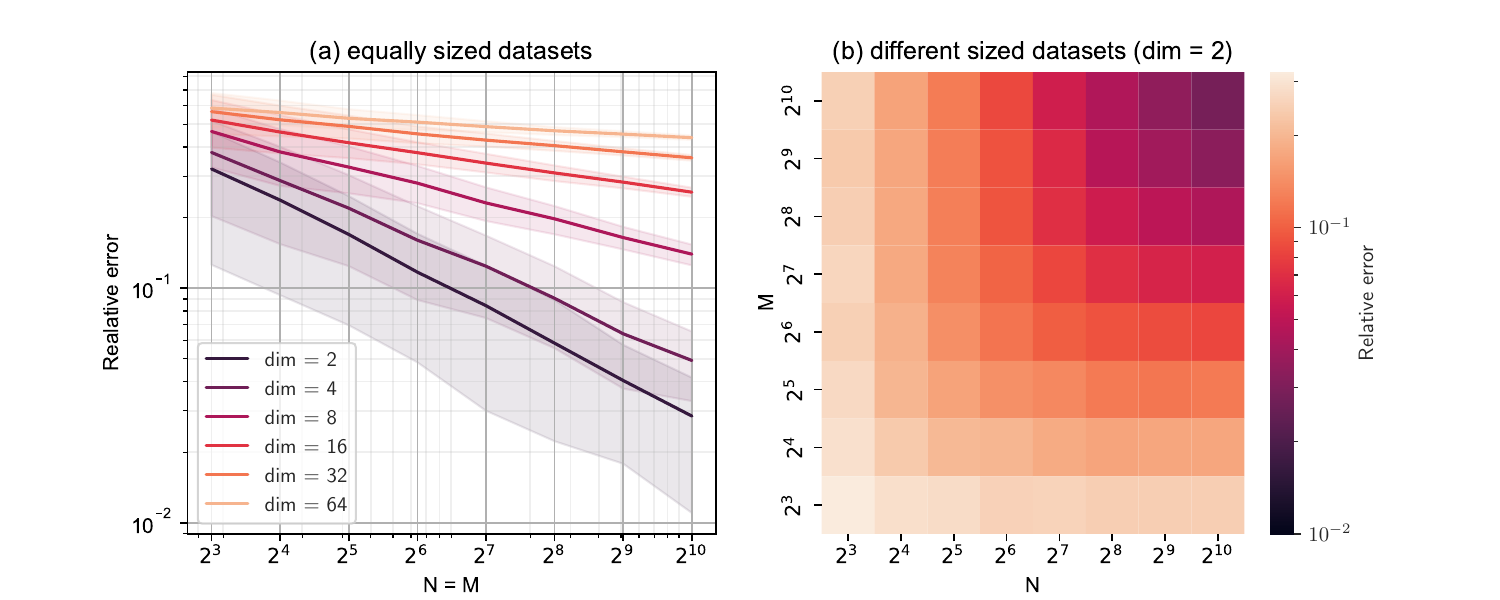}
	\caption{
		Error of the Wasserstein distance: We compare the numerical estimation of
		the Wasserstein distance between two data sets of cardinality N and M.
		The datasets are sampled from multivariate normal distributions of different dimensionality.
		They differ in their mean, i.e., a zero vector and a vector filled with ones, but share the covariance matrix $C_{ij} = \delta_{ij} + 0.2 \cdot (1 - \delta_{ij})$.
		For each combination (N, M), we sample 1000 datasets.
		The confidence interval in (a) highlights the area between the first and third quartiles.
		Numerical computation follows~\citep{Flamary2021pot, Dutta2021, Bonneel2011}, the analytic solution was discovered by~\citet{Dowson1982}.
	}
	\label{fig:wasserstein-gaussian}
\end{figure}

We emphasize that ABC only yields an approximation of the actual posterior distribution since it involves several assumptions. First, the bound $\epsilon$ appearing in the distance criterion \eqref{eq:abc_distance_criterion} introduces an approximation: if $\epsilon = 0$, the ABC algorithms would sample from the true posterior; however, for $\epsilon > 0$, the algorithms draw samples from an approximation to the true posterior. Instead of reaching a desired target $\epsilon$, the algorithm is often stopped after a fixed amount of dataset simulations contributing to the same error category. Second, the use of summary statistics introduces another level of approximation. This also holds for statistical distances; for instance, we need a sufficient number of data points in both sets to accurately estimate the Wasserstein distance between two distributions. Figure~\ref{fig:wasserstein-gaussian} illustrates this problem by displaying the relative error $ | \mathcal{W}_\mathit{true} - \mathcal{W}_\mathit{num} | / \mathcal{W}_\mathit{true} $ of the numerically computed Wasserstein distance for two multivariate Gaussian distributions. Lastly, the approximation of the posterior in terms of particles and kernel choice may affect the approximation~\citep{Sisson2019}.

\subsection{Software implementation and interfaces}\label{methods:software}

We leverage the implementation of the Del Moral's and Bernton's SMCABC algorithm~\citep{DelMoral2006, DelMoral2012, Bernton2019} provided in the ABCpy python package~\citep{Dutta2021}. Propagating the particles is the most expensive step of the algorithm because it involves the simulation of data under the model, i.e., at least $N \cdot \nsamples$ model evaluations per iteration for $N$ particles and $\nsamples$ samples per parameter. In the overview given in Figure~\ref{fig:project-overview}, this step is labeled with~\cnum{black}{white}{3}. However, the algorithm is inherently parallel in the particle updates, and the implementation offers parallel backends via Spark and MPI, of which we decided to use the latter. For code availability, we refer to the \nameref{suppl}.

In order to maximize the utilization of parallel computing resources, it is best practice to parallelize the outer loops and optimize the repeatedly executed code. Our implementation follows this logic by allowing ABCpy to parallelize the particle updates via MPI and implementing the ABM models with the highly efficient BioDynaMo~\citep{Breitwieser2021, Breitwieser2023} C++ simulation platform.

\begin{figure}
	\centering
	\includegraphics[width=0.9\textwidth]{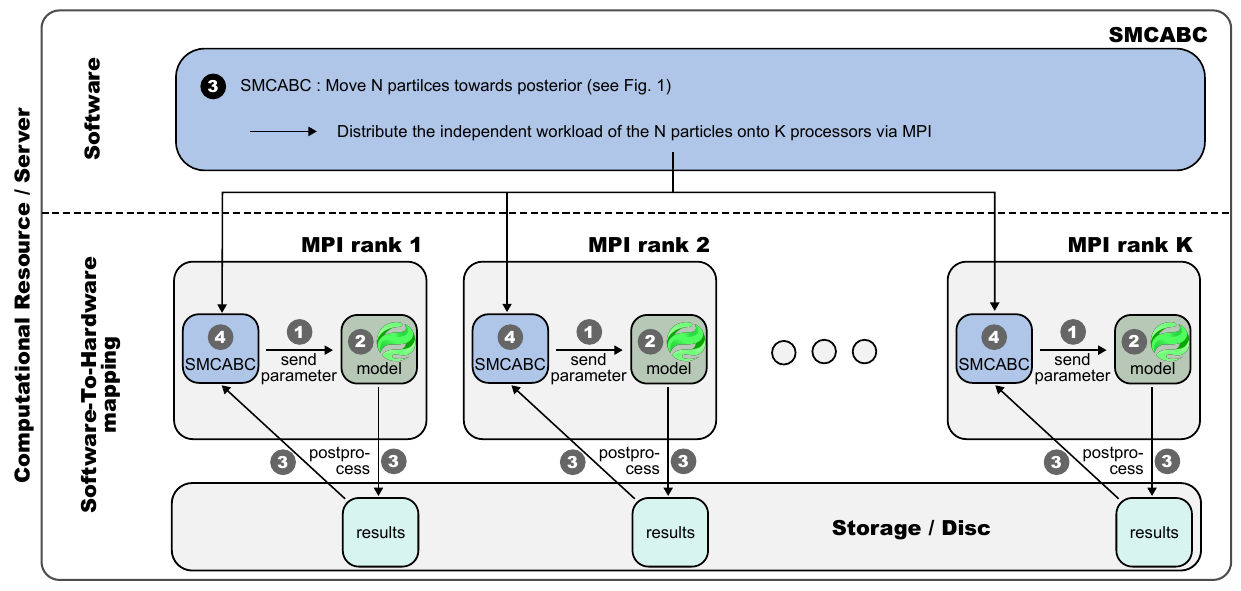}
	\caption{
		Software implementation and interfaces for the SMCABC sampling
		step~\cnum{black}{white}{3} in Figure~\ref{fig:project-overview}.
		Persistent BioDynaMo processes (green) reside on all server cores during calibration.
		The \textit{simulation objects} that ABCpy (blue) dynamically generates on the MPI ranks are connected to the respective BioDynaMo process via stdin-stdout.
		Results are written to and retrieved from a disk allocated in RAM to enhance performance.
		\cnum{gray}{white}{1} parameter proposal,
		\cnum{gray}{white}{2} simulate data under the model,
		\cnum{gray}{white}{3} compute morphometrics, and
		\cnum{gray}{white}{4} evaluate acceptance criterion.
	}
	\label{fig:software-implementation}
\end{figure}

During our initial computational experiments, we discovered a bottleneck in the current BioDynaMo version. Most of the execution time was spent starting the BioDynaMo simulation engine, more precisely, initializing the C++ interpreter \textit{cling}~\citep{Vasilev2012}. Initialization became the dominant factor since the simulation of a single neuron is very fast compared to the extensive simulations with billions of agents that BioDynaMo supports. To mitigate this performance bottleneck, we avoid launching a new simulation process for each parameter set. Instead, we start a persistent BioDynaMo simulation process for each MPI rank at the beginning of the calibration (here, one per core), keep it alive, and exchange data with ABCpy using \textit{IO-redirection}. Figure~\ref{fig:software-implementation} summarizes this concept and the implemented interface between the ABCpy and BioDynaMo software packages.

As ABCpy distributes the computation between the workers, a custom class facilitates the communication and data exchange with the active BioDynaMo processes. If a given SMCABC-MPI rank requires a model run for a set of parameters, this controller class requests the simulation from the BioDynaMo process running on the same core and waits until the simulation has been completed. The process writes the results, i.e., the SWC file of the synthetic neuron and possibly the associated morphometrics, to a RAM disk. After the simulation, the controller allows ABCpy to proceed - it retrieves the results, possibly applies some post-processing, and evaluates if the proposal parameters are accepted. Avoiding the repeated startup overhead, we measured 10x speedup compared to its initial version for simple models. We note that the more expensive the model, the less the coupling affects the runtime.

The implementation of the Sobol SA is analogous; here, the parameters are known a priori such that we assign different parameter combinations to different ranks and process them one after another using the same interface. The initial parameters are obtained from the python library SALib~\citep{SALib}.

\section{Results}

In this section, we explore the numerical experiments and their results conducted to foster an understanding of the stochastic models and the inverse problem. We first investigate the models' stochasticity to understand how the stochastic components of the model influence the QoIs for a fixed parameter choice. With a Sobol SA, we allow the model parameters to vary and explore how the variations affect the model outputs. We then study the stochastic inverse problem for the mechanistic neuronal growth Model~2 for synthetic data; in particular, we investigate how different choices of morphometrics, statistical distances, and simulated data set sizes affect our ability to recover the data-generating parameters. Afterward, we treat experimental data and calibrate Model~1 and~2 such that they mimic pyramidal cells in the human and mouse hippocampus (CA1 region) and extend our analysis beyond the QoIs with 3D visualizations. Lastly, we comment on runtime and computational costs.

\subsection{Model stochasticity}

We analyze the influence of the stochastic model components on the QoIs (morphometrics). We choose a fixed parameter vector $\Theta^\star$ for each model and simulate different neurons (i.e., same model, same parameter, different random seed). Table~\ref{tab:parameter-stochasticity-study} in Appendix~\ref{ap:parameter} summarizes the parameter choice for each model and are taken from~\citet{Breitwieser2021}. We generate $10^4$ artificial neurons per model and compute the QoIs. We then determine and visualize each model's (marginal) distribution of the QoIs.

\begin{figure}
	\includegraphics[width=\textwidth]{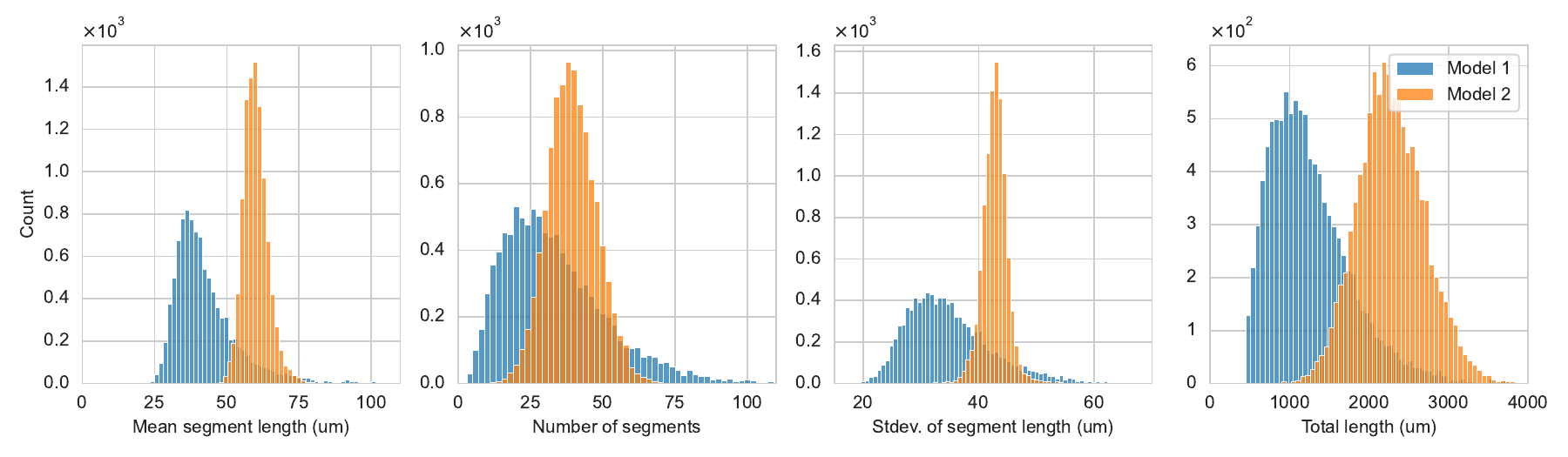}
	\caption{
		Simulated data sets simulated under Model~1 (blue) and Model~2 (orange).
		The histograms are computed on $10^4$ simulated neurons.
	}
	\label{fig:model-stochasticity}
\end{figure}

Figure~\ref{fig:model-stochasticity} shows histograms of the mean and standard deviation of the segment lengths together with the number of segments and total length. Across all the QoIs, Model~1 shows distributions that are significantly wider, less symmetric, and have a heavier tail compared to the output of Model~2. The distributions of Model~2 are, in most cases, symmetric and centered. The marginals of Model~2 appear to be similar to the characteristic shape of the normal distribution; nonetheless, the Anderson-Darling test allows us to reject the hypothesis that Model~2's marginals follow a normal distribution at a significance level of 1\%. Table~\ref{tab:descriptive_statistics_syndata} in Appendix~\ref{ap:parameter} additionally depicts the descriptive statistics of the marginals depicted in Figure~\ref{fig:model-stochasticity}. These statistics underline and quantify the previous observations, for instance, the apparently wider distributions of Model~1 reflect in a larger standard deviation. We remark that although the models are similar and mainly differ in how they distribute resources while branching, the QoI marginals show significant differences. Moreover, the presented data shows that the neuron models are truly stochastic and they must consequently be treated in a probabilistic setting.

\subsection{Sensitivity analysis}

We continue analyzing the behavior of the forward models, now striving to understand if some parameter influences a given QoI and, if so, by how much. Therefore, we probe the models with a global Sobol SA~\citep{Sobol2001, Saltelli2002, Saltelli2010}. We define each model's parameter domain $\Omega \in \mathbb{R}^n$ according to Table~\ref{tab:parameter-sensitivity-study} in Appendix~\ref{ap:parameter}.

\begin{figure}
	\includegraphics[width=\textwidth]{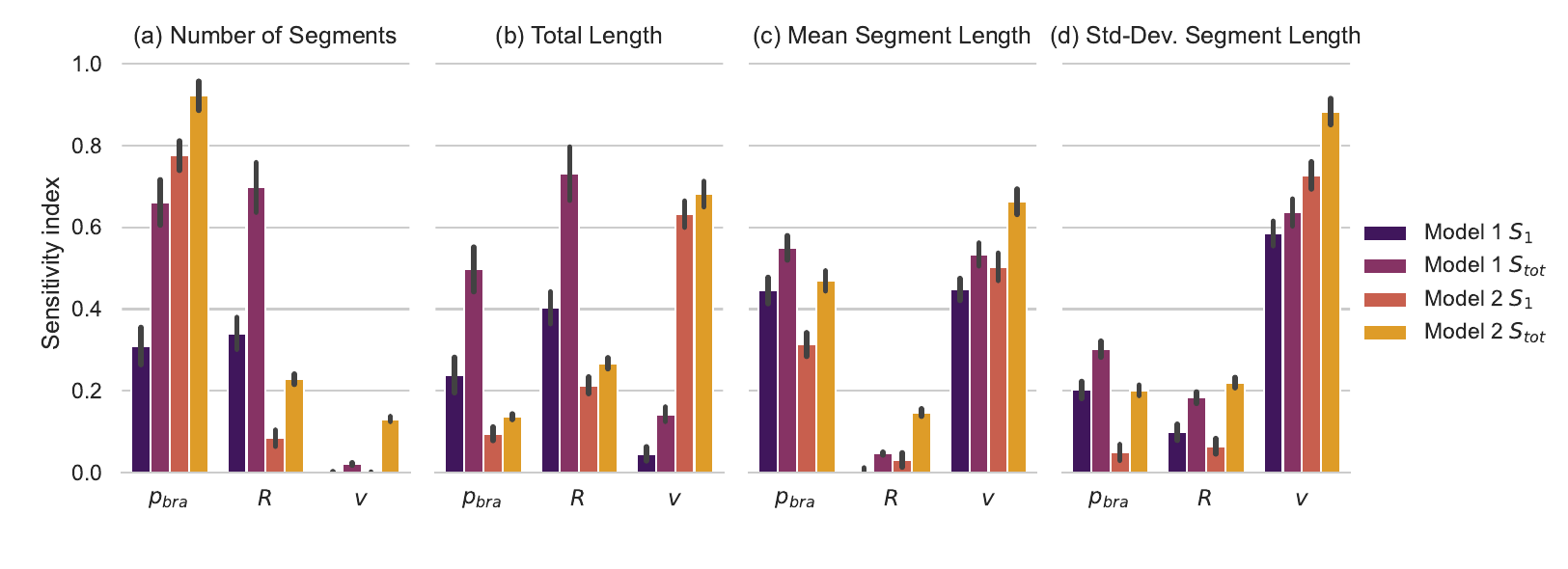}
	\caption{First order and total sensitivity indices of the growth models. The vertical gray line indicates the 95\% confidence interval of the index. The confidence interval does not account for the Monte Carlo error of $\mathbb{E}[QoI]$.}
	\label{fig:sensitivity}
\end{figure}

For each model, Saltelli sampling results in $K=N \cdot (2 \cdot \dim (\Omega) + 2) = 32768$ parameter combinations for which we need to evaluate the model (we chose $N=4096$). To account for the stochasticity of the model, we simulate $\nsamples=20$ samples for each parameter combination to estimate the expectation value of the different QoIs. Thus, we evaluate $6.6 \cdot 10^5$ artificial neurons per model to measure the first oder and total sensitivity indices ($S_1$, $S_{tot}$). We analyze the influence of the branching probability~($p_{bra}$), the resource consumption~($R$), and the elongation speed~($v$) on the QoIs. We restrict ourselves to this set of parameter for two reasons. First, simple experiments with the models showed that these parameter have a profound impact on the generated morphology and, second, the models differ in the resource reallocation while branching but tip agents in both models execute the same persistent biased walk. We thus decided to focus on the parameter that differ between the models. The results are displayed in~Figure~\ref{fig:sensitivity}.

We begin with discussing the similarities highlighted by the sensitivity analysis. Figure~\ref{fig:sensitivity}~(c) shows that the mean segment length of both models reacts similarly to the parameters. For both models, this QoI is primarily influenced by $p_{bra}$ and $v$ while $R$ has no influence. This can be understood from realizing that the mean segment length is the average distance that a tip agent migrates between two branching events. The parameter $v$ is the speed of the tip migration while $p_{bar}$ defines the likelihood of branching per time unit. The resource consumption does not influence this pattern and, thus, the corresponding sensitivity indices are very low. Both models share the mathematical description of the tip migration which reflects in the sensitivity analysis. This argument extends to the standard deviation of the segment length.

In contrast to the mean and standard deviation of the segment length, the number of segments and the total length in Figure~\ref{fig:sensitivity}~(a,b) respond differently to the parameter variations. The analysis shows that Model~1 is much more sensitive to the parameter $R$. Considering the regime for very low values of $R$, Model~1 can repeatedly split newly created branches leading to an exponential growth in the number of branches. In the same scenario, Model~2 remains well controlled and continues to produce a main brach with short side extensions. This realizations explains why Model~1 is more sensitive to $R$. For Model~2, the number of segments is primarily influenced by $p_{bra}$. Similar arguments hold for the total length, here the exponential growth outweighs the linear growth parameter $v$ for Model~1 while Model~2 is mostly sensitive to it. We further observe that Model~1 shows statistically significant gaps between the first order and total sensitivity index, hinting at higher order effects. In its sum, these observations underline the significant impact that the resource distribution has on the morphology.

\subsection{Solving the stochastic inverse problem with SMCABC}

We study the stochastic inverse problem arising for mechanistic neuronal growth models with the SMCABC algorithms altering the statistical distance measures, algorithmic parameters, morphometrics, and data sets. We first treat synthetic data to verify that we may successfully recover the data-generating parameter in a setting where the model can reproduce the data well. The synthetic data sets used in the calibration are subsets of the ones in the section analyzing the models' stochasticity. We perform extensive numerical experiments to showcase how the calibration algorithm behaves under different circumstances. Afterward, we apply the SMCABC algorithm to the data sets defined in Table~\ref{tab:overview-data} to determine which models adequately describe given experimental data. The majority of the computational experiments deal with Model~2 because it models the structurally more complex part of the neuron (i.e., top part in Figure~\ref{fig:neuron-abm}).

\subsubsection{Synthetic data}

We begin with calibrating Model~2 with synthetic data comprised of 500 synthetic neurons. We attempt to find the posterior distribution for the same model parameters ($p_{bra}$, $R$, and $v$) considered in the SA. The synthetic data was generated with the parameter choice $p_{bra}^\star = 0.38 \cdot 10^{-1}$, $R^\star = 0.71 \cdot 10^{-3}$, and $v^\star = 10^{2}$; recovering these parameter serves as test case throughout this section. We employ the morphometrics to map the simulated neurons to $\mathbb{R}^n$, effectively reducing a neuron morphology to a $n$-dimensional random variable ($n=1,\dots,4$). We provide uniform priors for all parameters, use $2^{10}$ particles to approximate the posterior and run the algorithm for a fixed budget of $5\cdot10^7$ growth-model simulations. We do not interrupt the algorithm in between SMC iterates; we either automatically stop the calibration after the iteration that has exhausted the simulation budget, or if the algorithm's runtime surpasses a threshold $T_{max}$.

\paragraph{Effect of the morphometrics.}

\begin{figure}
	\centering
	\subfloat[mean of segment lengths (SLs) ($\mathcal{M}2$)]{
		\includegraphics[width=\textwidth]{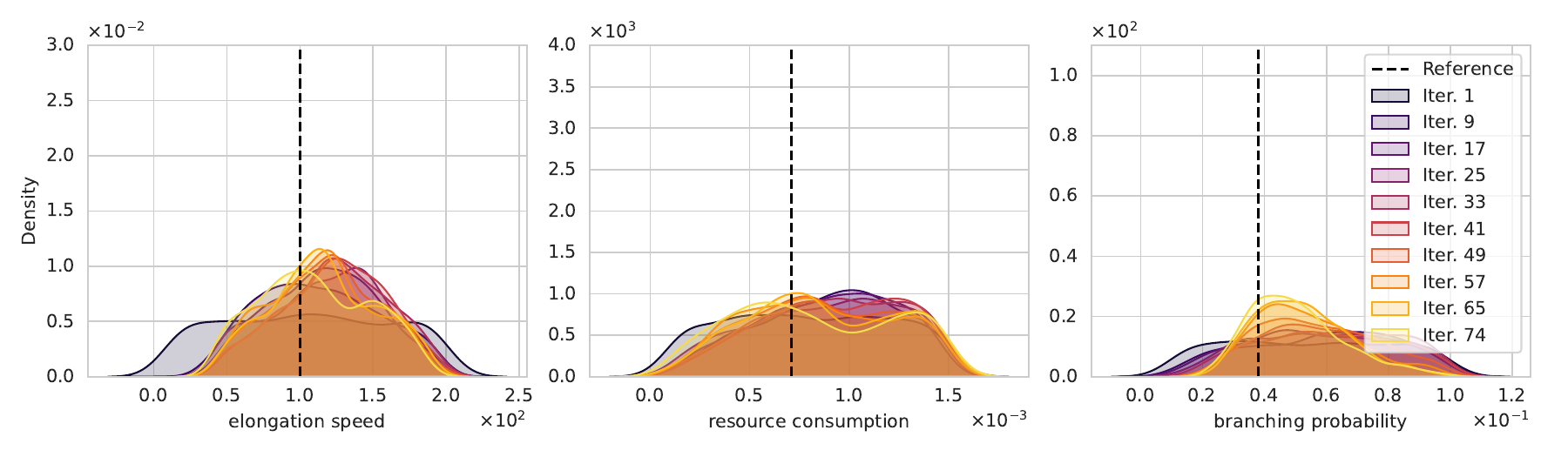}
	} \\
	\subfloat[mean and standard deviation of SLs ($\mathcal{M}2,3$)]{
		\includegraphics[width=\textwidth]{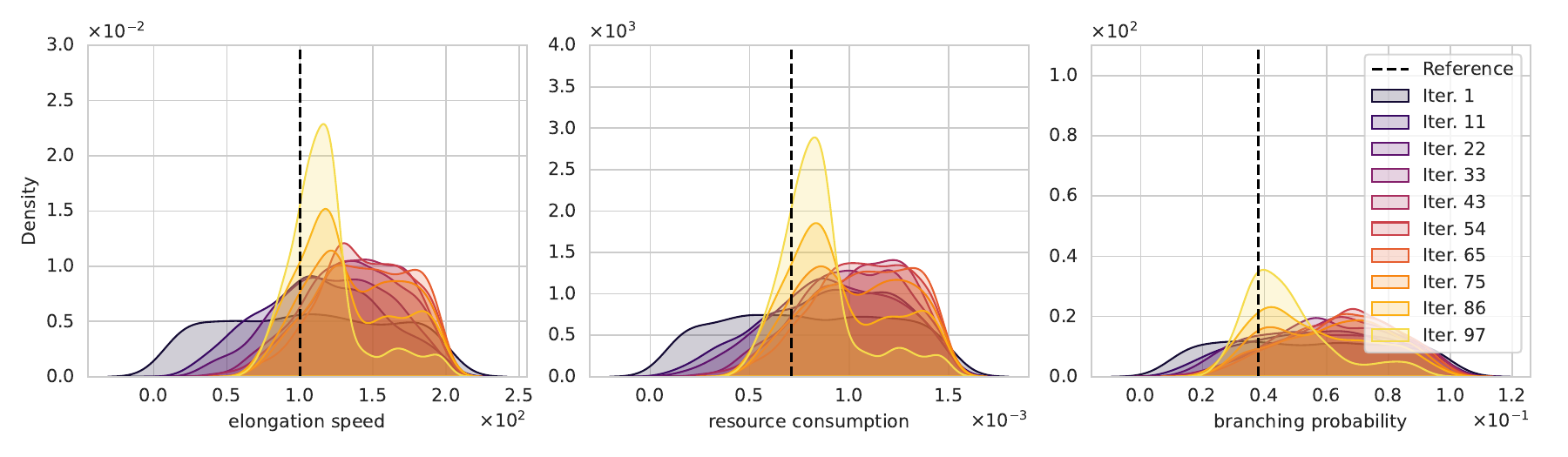}
	} \\
	\subfloat[mean and standard deviation of SLs, number of segments ($\mathcal{M}1,2,3$)]{
		\includegraphics[width=\textwidth]{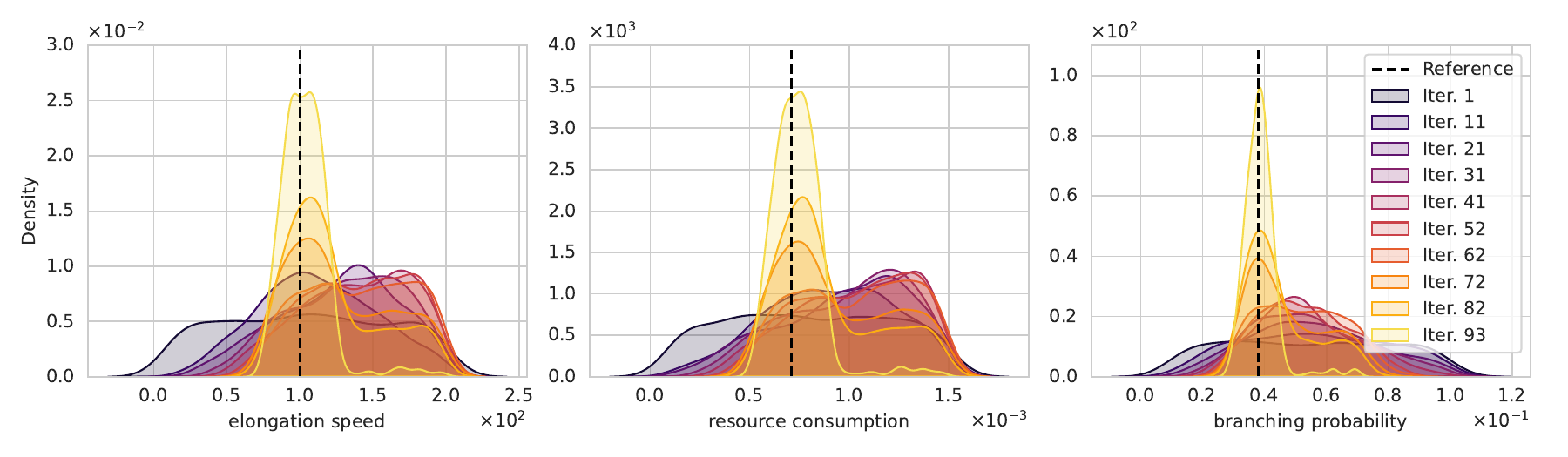}
	} \\
	\subfloat[mean and standard deviation of SLs, number of segments, total lenth ($\mathcal{M}1,2,3,4$)]{
		\includegraphics[width=\textwidth]{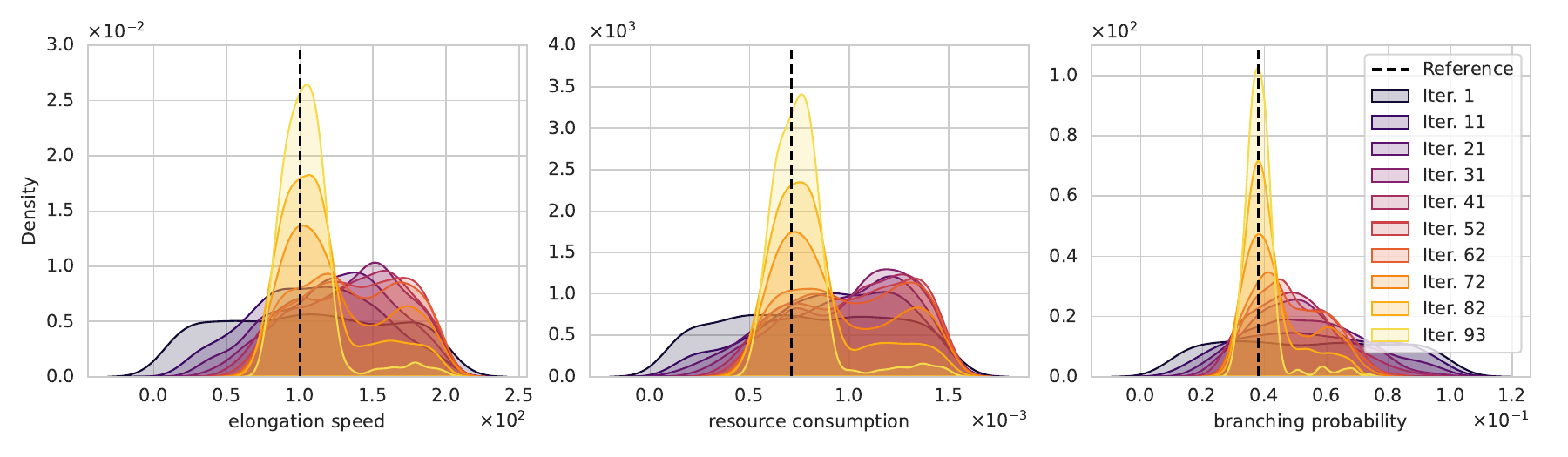}
	}
	\caption{
		Convergence of the Wasserstein posterior under different choices of
		morphometrics.
		The information captured by the morphometrics increases from~(a) to~(d), and the posterior improves accordingly.
	}
	\label{fig:smcabc-convergence-effect-morphometrics}
\end{figure}

First, we investigate the effect of the choice of morphometrics. Intuitively, they capture the information of the neuron morphology. If the morphometrics capture insufficient information, the algorithm will not be able to recover the data-generating parameter for the simple reason that the information needed for their inference is not considered. Thus, the first experiment explores the convergence of the posterior for different combinations of morphometrics, specifically $\mathcal{M}$1, 2, 3, and 4, i.e., the number of segments, the mean and standard deviation of the segment lengths, and the total length. This experiments uses the Wasserstein distance to evaluate~\eqref{eq:abc_distance_criterion} and $\nsamples = 50$ samples per parameter.

Figure~\ref{fig:smcabc-convergence-effect-morphometrics} shows how different morphometrics choices affect the posterior marginals. From~(a) to (d), we add one dimension to the morphometrics at a time. We begin with the mean of the segment lengths~(a) and add the standard deviation~(b), the number of segments~(c), and, lastly, the total length~(d). All four plots depict the evolution of the posterior marginals over the SMC iterations. The marginals are \textit{Gaussian kernel density estimates} (KDEs) computed from the particles and associated weights. Early iterations appear dark (black, purple), and final iterations in bright colors (orange, yellow). For convenience, the data-generating parameters $p_{bra}^\star$, $R^\star$, and $v^\star$ are indicated with vertical, dashed, and black lines. We judge the algorithm performance by its ability to recover $p_{bra}^\star$, $R^\star$, and $v^\star$, i.e., good algorithmic setups are expected to show posterior marginals condensing around the vertical black lines. This type of visualization will reoccur in the other numerical experiments.

From Figure~\ref{fig:smcabc-convergence-effect-morphometrics}, we see that the mean segment length is not sufficient to recover the parameter. The posterior in~(a) is very wide indicating high parameter uncertainties. Adding the standard deviation of the segment length improves upon the previous case, we find sharper posterior marginals, however, their peaks show a slight offset incorrectly identifying the parameters $v$ and $R$. The parameter $p_{bra}$ centers around the correct value, however, the width of the posterior again indicates low confidence. Adding the number of branches to the QoIs significantly improves the marginals' quality; all three parameter peak at the data-generating parameter and are strongly centered indicating good confidence in the identified parameters. We emphasize the substantial improvement in identifying the branching probability which can be understood from the results of the SA; it demonstrated that the the number of branches of Model~2 is very informative concerning this parameter. Lastly, adding the total dendritic length seems to neither harm nor further improve the posterior.

To further verify the posterior quality, we measured the Wasserstein distance between a second dataset generated with identical parameters and the one used for calibration. The measured distance is in line with the final $\epsilon$ value supporting the claim that the algorithm found a good posterior distribution whose simulation are close to indistinguishable from the calibration data. While there are barely differences in the posterior quality of~(c) and~(d), we favour the latter for computational reasons. Using four instead of three morphometrics converged quicker and the overall runtime was roughly the half. For all subsequent experiments, we use the four morphometrics $\mathcal{M}1,2,3,4$.

\paragraph{Effect of different statistical distances.}

\begin{figure}
	\centering
	\subfloat[KL-divergence]{
		\includegraphics[width=\textwidth]{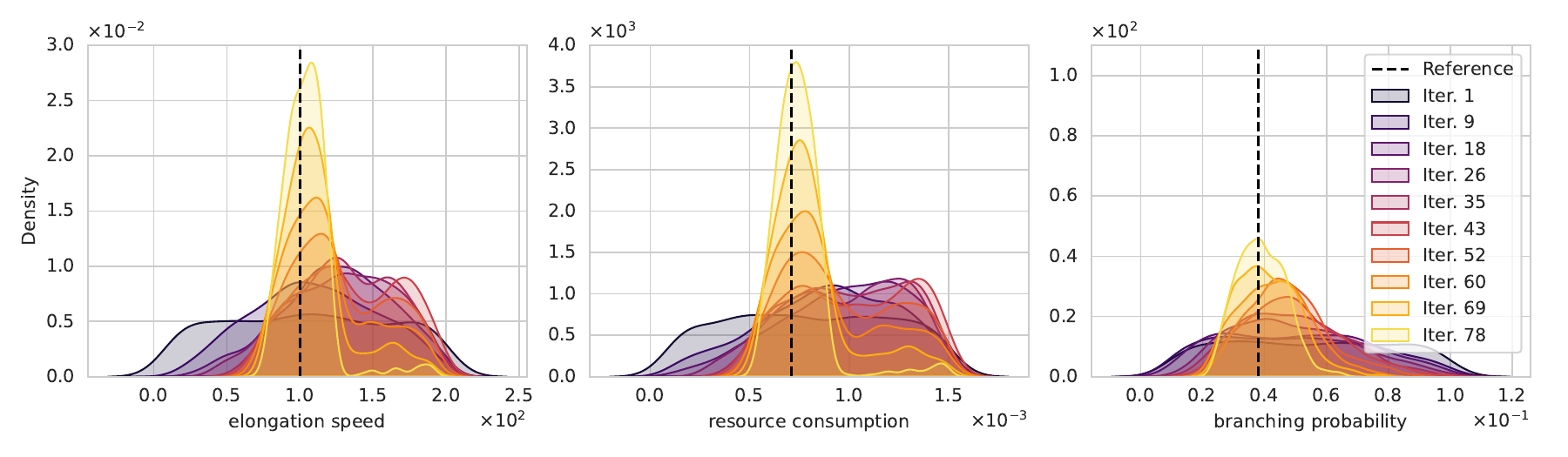}
	} \\
	\subfloat[Gamma-divergence]{
		\includegraphics[width=\textwidth]{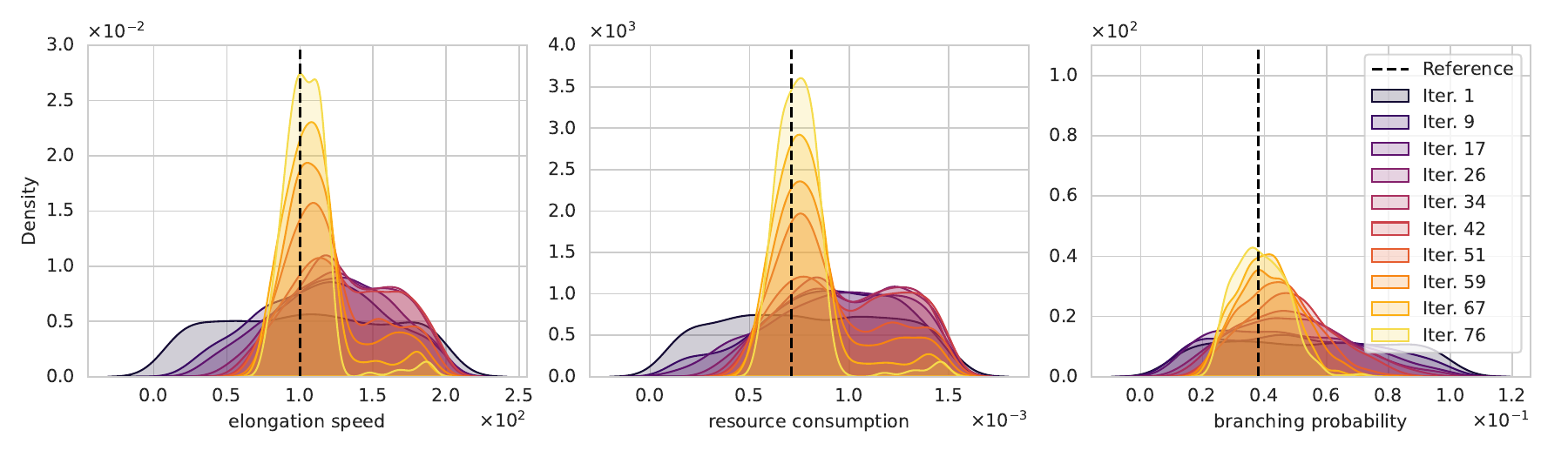}
	}
	\caption{
		Convergence of the ABC algorithms over SMC iterations for different
		distribution-based distance metrics.
		All algorithms can simulate $10^6$ datasets of cardinality $\nsamples=50$; however,~(a) and~(b) stalled after performing 82\% and 85\% of the allowed simulations, respectively.
		Wasserstein better identifies the branching probability, see Figure~\ref{fig:smcabc-convergence-effect-morphometrics}~(d).
	}
	\label{fig:smcabc-convergence-distance-comparison}
\end{figure}

We proceed with analyzing how different statistical distances (Wasserstein~\citep{Bernton2019}, sliced-Wasserstein~\citep{Kimia2020}, KL~\citep{Jiang2018}, $\gamma$~\citep{Fujisawa2021}) affect the calibration results. We run the SMCABC algorithm with the same parameters but vary the statistical distances. Figure~\ref{fig:smcabc-convergence-distance-comparison} displays our findings. The interpretation of the graphs follows the previous section; each plot shows the evolution of the posterior marginals over the SMC iterations with the data-generating parameter indicated by vertical, dashed, and black lines.

The experiment shows that the Wasserstein distance in Figure~\ref{fig:smcabc-convergence-effect-morphometrics}~(d) as well as the KL and $\gamma$-divergence in Figure~\ref{fig:smcabc-convergence-distance-comparison}~(a) and (b), respectively, recover the parameter well and find concentrated posterior distributions around the reference values. KL and $\gamma$ divergence seem to require slightly fewer SMC iterations to concentrate around the data-generating parameter of the elongation speed $v$ and resource consumption $R$. This can be seen from comparing Figure~\ref{fig:smcabc-convergence-effect-morphometrics}~(d, iteration 82) with Figure~\ref{fig:smcabc-convergence-distance-comparison}~(a, iteration 78) and (b, iteration 76), although this observation is unlikely to be statistically significant. Evidently, the Wasserstein distance outperformed the other distances in identifying the branching parameter.

Neither the KL divergence nor the $\gamma$ divergence exhausted the total number of dataset simulations. After they used 82\% and 85\% of their budget, respectively, individual particles got trapped in regions with low posterior probability, which caused a significant load imbalance, making it infeasible to let the algorithm run until the end. Similar problems were not observed for the Wasserstein-based inference, which reliably converged in all calibration runs. The sliced-Wasserstein distance failed to discover the data-generating parameter in this particular case (Fig.~\ref{fig:s1-calibration-sliced-wasserstein} in Appendix~\ref{sup:s1-calibration-sliced-wasserstein}). While the posterior marginals peak at the correct values, the posterior is significantly wider than for the other statistical distances.

\paragraph{Effect of simulated dataset size.}

\begin{figure}
	\centering
	\subfloat[$\nsamples=100$ samples per parameter]{
		\includegraphics[width=\textwidth]{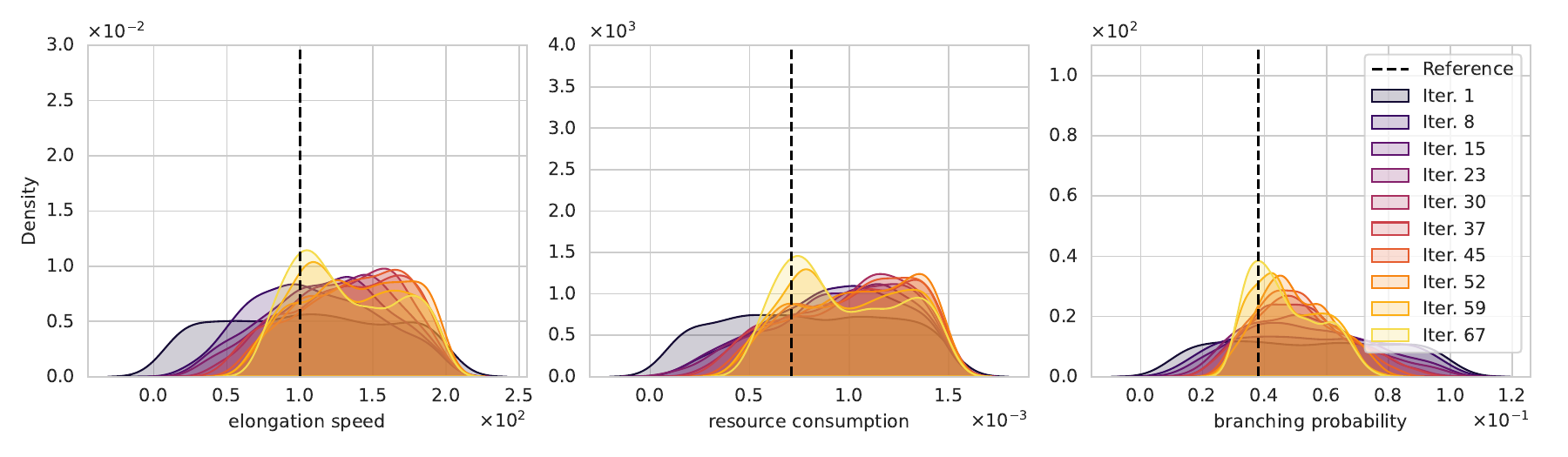}
	} \\
	\subfloat[$\nsamples=25$ samples per parameter]{
		\includegraphics[width=\textwidth]{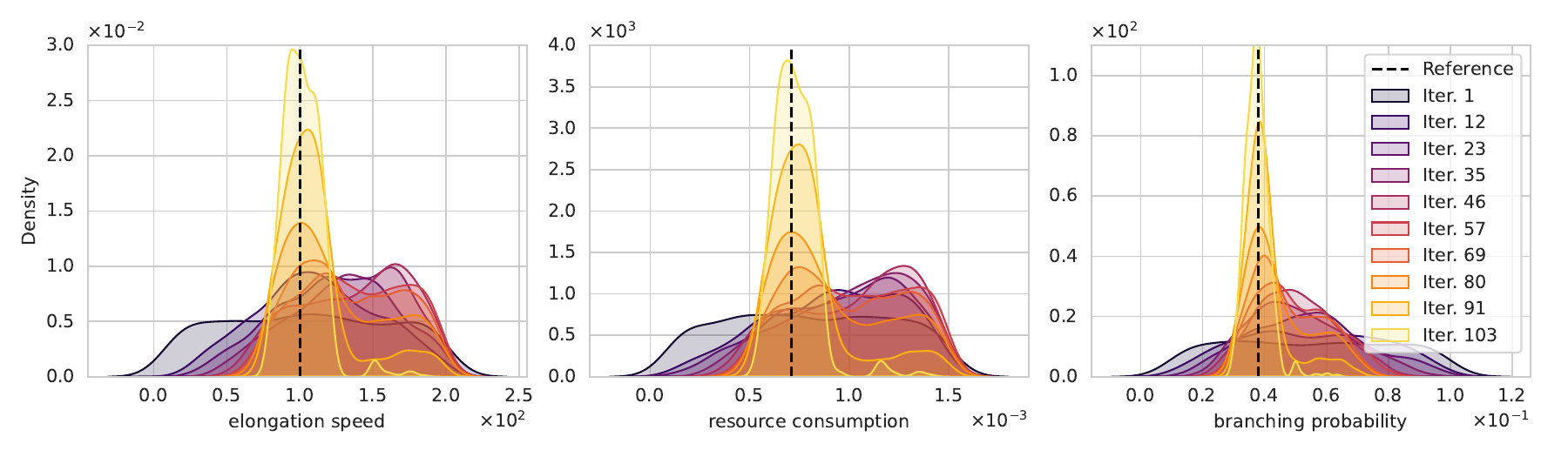}
	} \\
	\subfloat[$\nsamples=10$ samples per parameter]{
		\includegraphics[width=\textwidth]{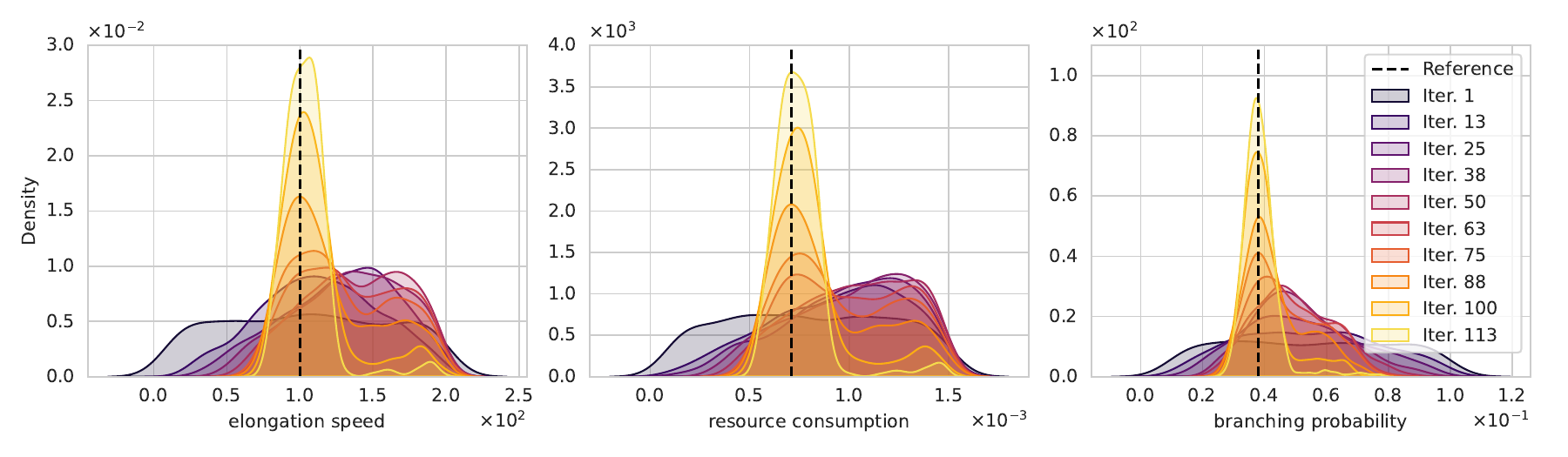}
	}
	\caption{
		Convergence of the Wasserstein ABC algorithms over SMC iterations for different
		sizes of the synthetic datasets.
		All algorithms simulate less than $5 \cdot 10^7$ neurons.
	}
	\label{fig:smcabc-convergence-effect-datasize}
\end{figure}

The previous experiments used $\nsamples=50$ synthetic neurons per parameter. This number was motivated by the original work on Wasserstein-ABC~\citep{Bernton2019} in which the authors used a sample size of $\nsamples=100$ to calibrate the mean vector of a bivariate Gaussian model. Since generating a neuron morphology is disproportionally more expensive, we began experiments with $\nsamples=50$ neurons and allowed the same number of dataset simulations ($10^6$) as~\citet{Bernton2019}. To reduce the algorithm's runtime, \citet{Bernton2019} considered less expensive approximations to the Wasserstein distance, e.g., Hilbert and swapping distances, because its computation accounted for large fractions of the overall runtime. Here, simulating 50 synthetic neurons is roughly 100 times more time-consuming than the distance computation; thus, we investigate how the sample size $\nsamples$ affects the posterior quality in this practical example. Allowing the same number of total neuron simulations ($5\cdot10^7$), we run the algorithm with $\nsamples \in \{10,25,50, 100\}$ samples per parameter while using $\mathcal{M}1,2,3,4$ and the Wasserstein distance.

Figure~\ref{fig:smcabc-convergence-effect-datasize} shows the evolution of the posterior marginals considering synthetic datasets $y_{\mathit{sim}}$ of cardinality~(a) 100,~(b) 25, and~(c) 10. Figure~\ref{fig:smcabc-convergence-effect-morphometrics}~(d) shows the identical experiment for 50 simulated neurons per parameter which recovers the data-generating parameter well. First, in Figure~\ref{fig:smcabc-convergence-effect-datasize}~(a), doubling the number of samples from 50 to 100 gives more confidence in the prediction of the model for a given set of parameters; however, the computational cost double and render the execution of many SMC iterations infeasible. After 67 iterations, the algorithm exhausts its simulation budget and yields insufficient posterior marginals to identify the data-generating parameters. Comparing similar iterations (62, $\nsamples=50$; 67, $\nsamples=100$), we may see a slight advantage for using more samples.

Next, we reduce the sample size to (b) 25 and (c) 10 per parameter. We expect that lowering the sample size allows the number of SMC iterations to increase. Simultaneously, the statistics of the model at a given parameters set are more uncertain; colloquially, we may say that the coupling between parameter and QoI space is looser. For both cases, the posterior marginals in Figure~\ref{fig:smcabc-convergence-effect-datasize} initially converge quickly. Towards the end, the sampling got inefficient and developed a load imbalance with individual particles holding up the algorithm. We eventually stopped the algorithm after it ran on the server as long as the reference results for 50 samples. To find a posterior distribution similar in quality, the algorithm with $n=25$ and $n=10$ samples used only 69\% and 41\% of the simulation budget, respectively. While 50 samples per parameter took roughly 4.5 days to recover the parameter,~(b) and~(c) achieved the same in roughly two days. Thus, the runtime is significantly reduced by lowering the sample size while still achieving a similar posterior quality. Additional benefits are not observed, i.e., the posterior is similar but does not show higher densities.

\subsubsection{Experimental data}

We now shift our attention towards experimental data. We begin with calibrating Model~2 on the apical dendrites of the data sets listed in Table~\ref{tab:overview-data} using the morphometrics $\mathcal{M}$1, 2, 3, and 4 (see Table~\ref{tab:overview-morphometrics}). The apical dendrites refers to the structures displayed above the soma in Figure~\ref{fig:neuron-abm}~(c). We approximate the posterior with $2^{10}$ particles, run the algorithm for a fixed budget of $5 \cdot 10^{7}$ neuron simulations ($10^6$ data sets $y_{\mathit{sim}}$ of cardinality 50), and use the Wasserstein distance. In contrast to using synthetic datasets, the reference parameters to judge the quality of the solution are unknown; hence, we fall back to the \textit{predictive check} described earlier, i.e., we sample $2^{10}$ parameter from the posterior, evaluate the model $\nsamples=50$ times per sample, compute the QoIs, and compare them with the data. Our focus lies on finding the posterior distribution for a given model -- whether or not the model adequately describes the data is beyond the scope of the present article and designing rules for specific resource-driven neuron models for given data is left for future work.

\begin{figure}
	\centering
	\subfloat[Posterior marginals]{
		\includegraphics[width=\textwidth]{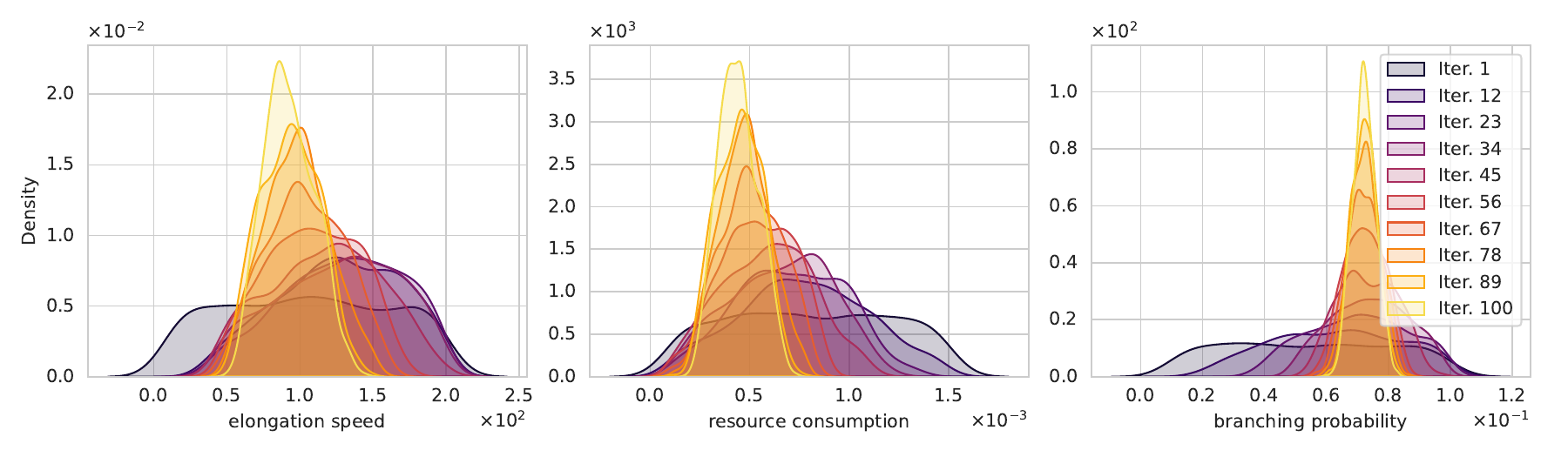}
	} \\
	\subfloat[Predictive check]{
		\includegraphics[width=\textwidth]{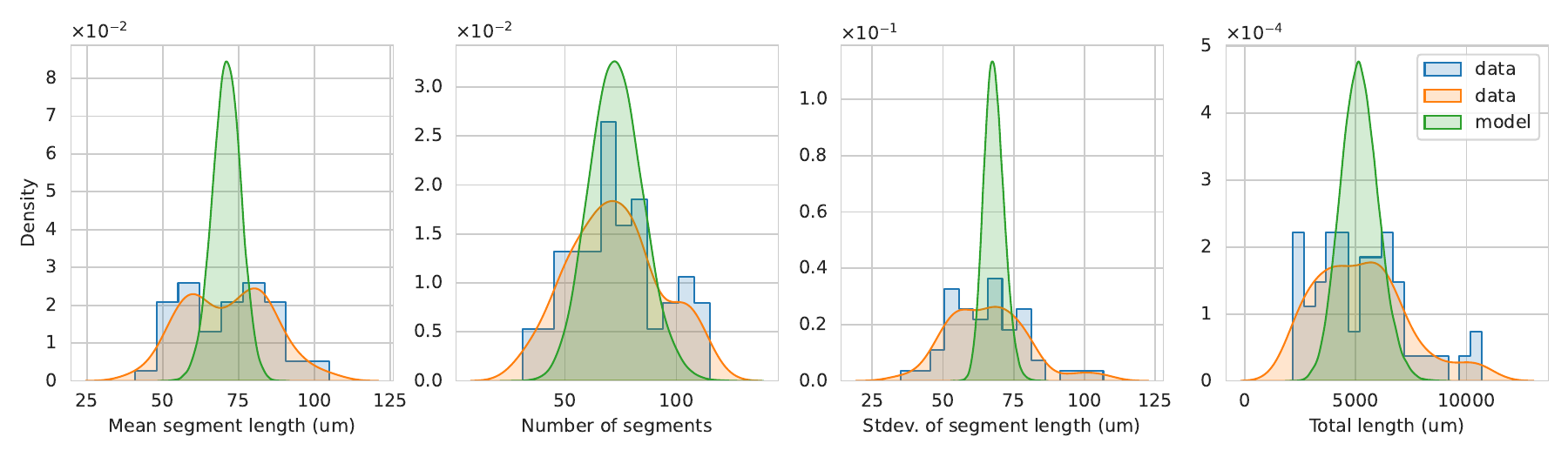}
	}
	\caption{
		(a) Convergence of Wasserstein posterior marginals and (b) predictive check of Model~2's calibration on experimental data $\mathcal{D}1$ (human, apical).
		The predictive check shows simulated samples drawn from the posterior distribution (green KDE) together with the data obtained from neuromorpho.org (blue histogram, orange KDE).
	}
	\label{fig:smcabc-convergence-experimental-data-1-apical}
\end{figure}

\paragraph{$\mathcal{D}1$ - pyramidal cells in human hippocampus (CA1).}
Figure~\ref{fig:smcabc-convergence-experimental-data-1-apical} shows the result of calibrating Model~2 with data $\mathcal{D}$1.
Panel~(a) shows the familiar evolution of the posterior marginals over the SMC iterations.
The marginals show a convergent behavior and accumulate in certain regions, specifically the KDEs are centered around $v^\star \approx 0.83 \cdot 10^{2}$, $R^\star \approx 0.44 \cdot 10^{-3}$, and $p_{bra}^\star \approx 0.72 \cdot 10^{-1}$.
Panel~(b) shows the results of the predictive check -- data in blue (histogram) and orange (KDE), model predictions in green (KDE).
The center (mean) of the marginal distributions of data and predictions match well.
The number of segments and the total segment length seem to behave similar for the experiment and the calibrated simulation.
The mean and standard deviation centers correctly but the model's prediction are too narrow indicating that additions to Model~2 may be necessary to better describe the data.

\begin{figure}
	\centering
	\subfloat[Posterior marginals]{
		\includegraphics[width=\textwidth]{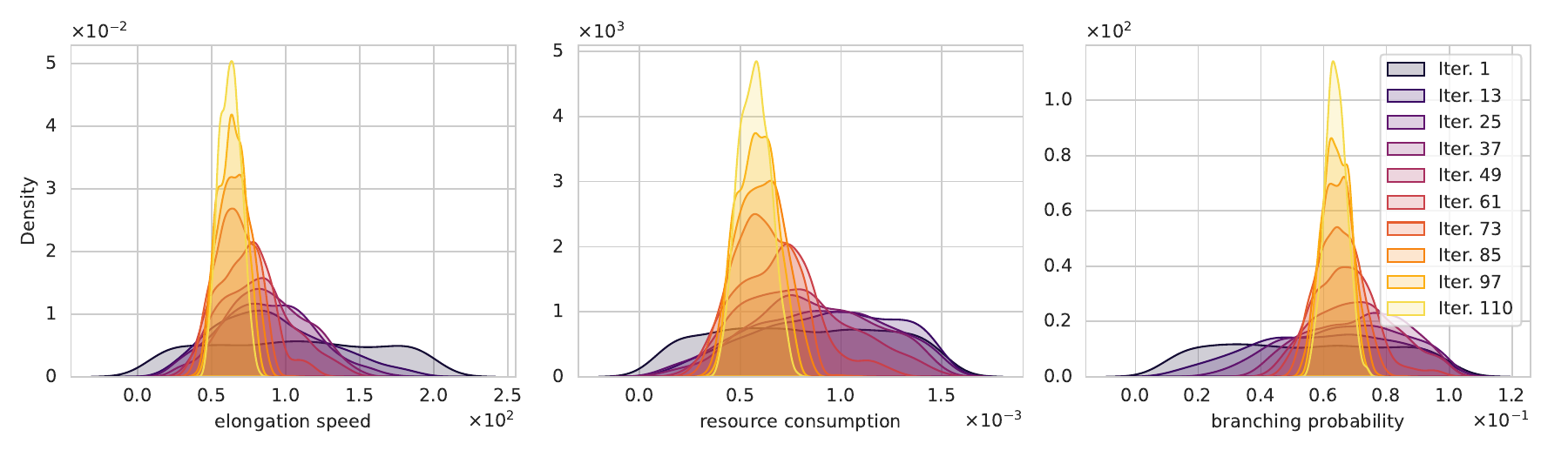}
	} \\
	\subfloat[Predictive check]{
		\includegraphics[width=\textwidth]{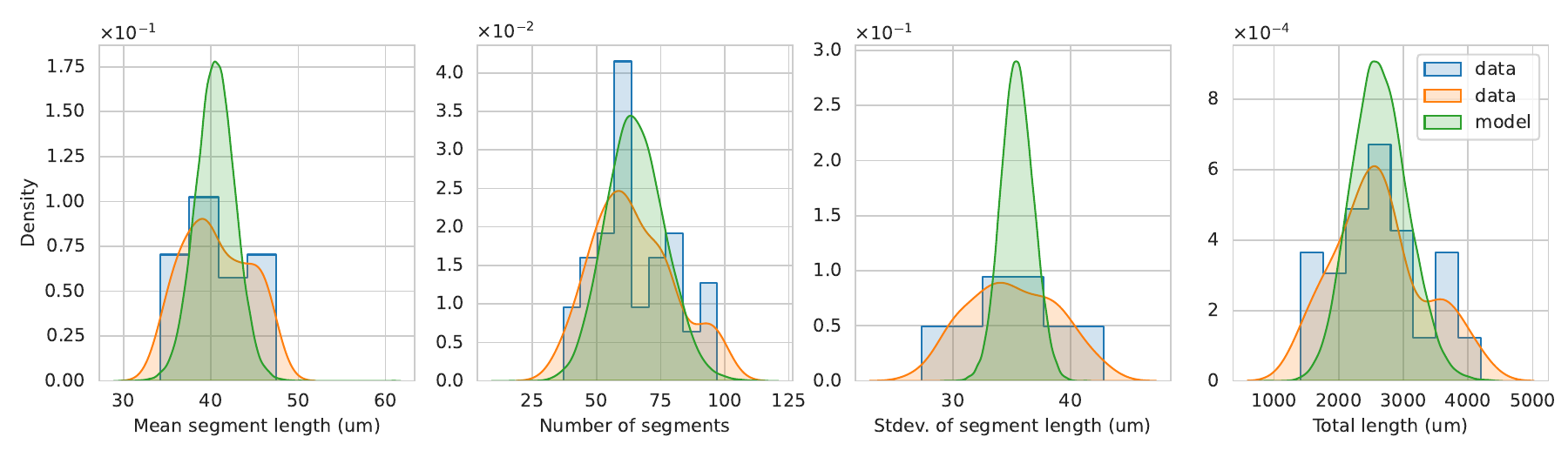}
	}
	\caption{
		(a) Convergence of Wasserstein posterior marginals and (b) predictive check of Model~2's calibration on experimental data $\mathcal{D}2$ (mouse, apical).
		The predictive check shows simulated samples drawn from the posterior distribution (green KDE) together with the data obtained from neuromorpho.org (blue histogram, orange KDE).
	}
	\label{fig:smcabc-convergence-experimental-data-2-apical}
\end{figure}

\paragraph{$\mathcal{D}2$ - pyramidal cells in mouse hippocampus (CA1).}
Figure~\ref{fig:smcabc-convergence-experimental-data-2-apical} shows the result of calibrating Model~2 with data $\mathcal{D}$2.
The interpretation is analogous to Figure~\ref{fig:smcabc-convergence-experimental-data-1-apical}.
In contrast to $\mathcal{D}1$, we obtain posterior marginals more concentrated around specific values, which can be seen from the higher density values of marginals.
In particular, these densities are higher for the elongation speed and the resource consumption.
Moreover, the shape of the  QoI marginals appears to match the shape of the data better compared to $\mathcal{D}1$.
The width of the standard deviation is, again, underestimated by the model.
For $\mathcal{D}2$, the marginals marginals peak at $v^\star \approx 0.58 \cdot 10^{2}$, $R^\star \approx 0.59 \cdot 10^{-3}$, and $p_{bra}^\star \approx 0.62 \cdot 10^{-1}$.
It appears as if Model~2 is better suited to describe pyramidal cells found in mice than the ones found in humans.

\paragraph{Comparison beyond the QoIs}

The selected QoIs ($\mathcal{M}1,2,3,4$) reduce the neuron morphology to a point in $\mathbb{R}^4$, representing a significant, irreversible compression of the information. To further investigate if Models~1 and~2 can produce meaningful, synthetic neurons, we visualize a set of neurons after calibration together with the calibration data.

We proceed as follows. First, we use Model~1 and~2 to describe the pyramidal cell's basal and apical parts, respectively, akin to the concept presented in Figure~\ref{fig:neuron-abm}~(b). Using the Wasserstein~ABC algorithm, we then calibrate both models as in the previous section. The calibration of Model~1 follows the calibration of Model~2 with datasets extracted from the basal structures. We select a parameter set from the resulting posterior marginals via the maximum likelihood paradigm and subsequently simulate $10^3$ neurons using the fixed parameter set. Compared to Figure~\ref{fig:neuron-abm}~(b), we lower the bias and increased the persistence of the random walk governing the growth. Lastly, we collect the basal and apical QoIs of the data and simulations, normalize them concerning the variance of the data marginals, and obtain two point clouds in $\mathbb{R}^8$. We assign a simulation to each data point via the shortest Euclidean distance to form data-simulation pairs.

\begin{figure}
	\centering
	\subfloat[neuron-id 1]{
		\includegraphics[width=0.33\textwidth]{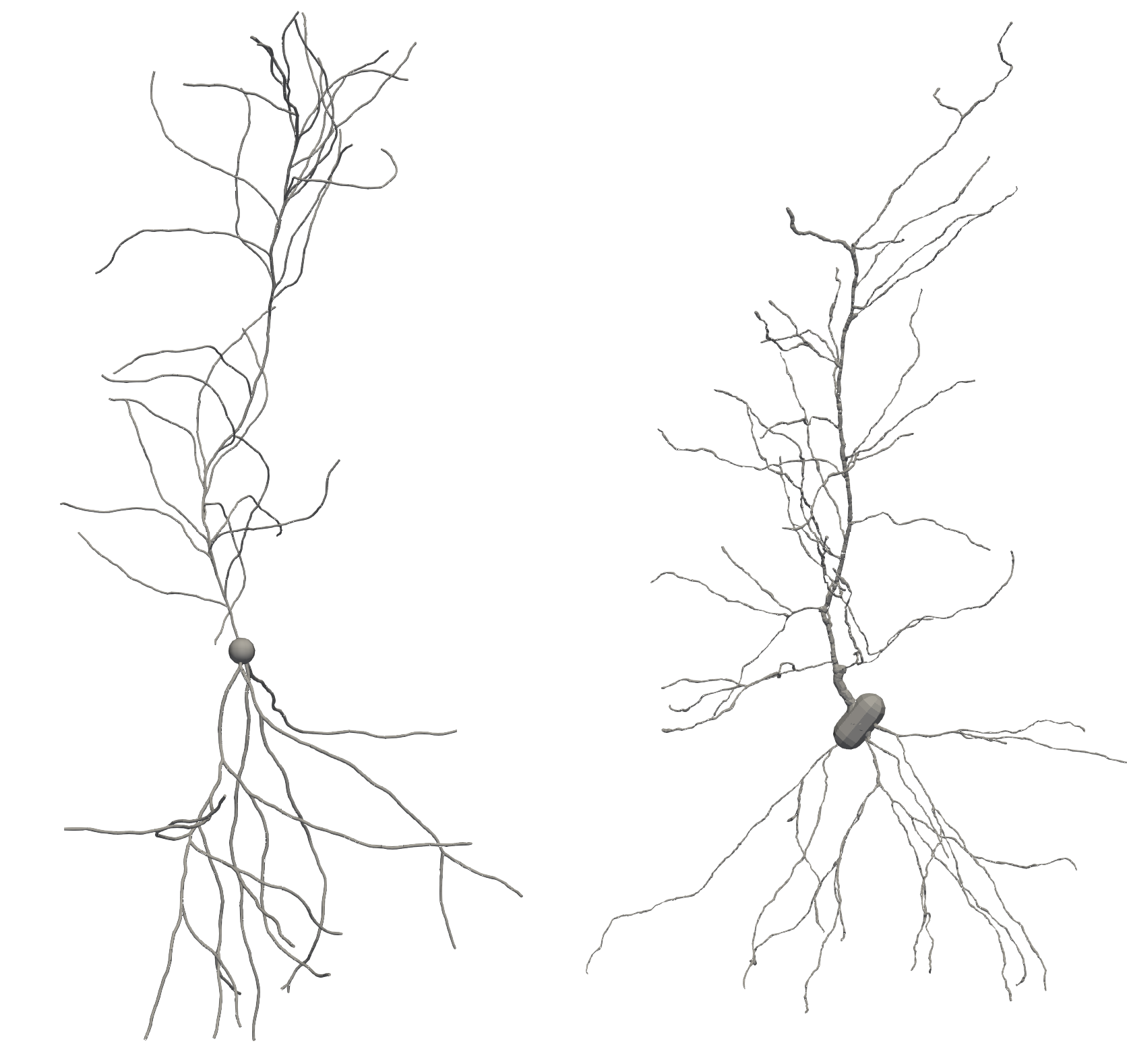}
	}
	\subfloat[neuron-id 6]{
		\includegraphics[width=0.33\textwidth]{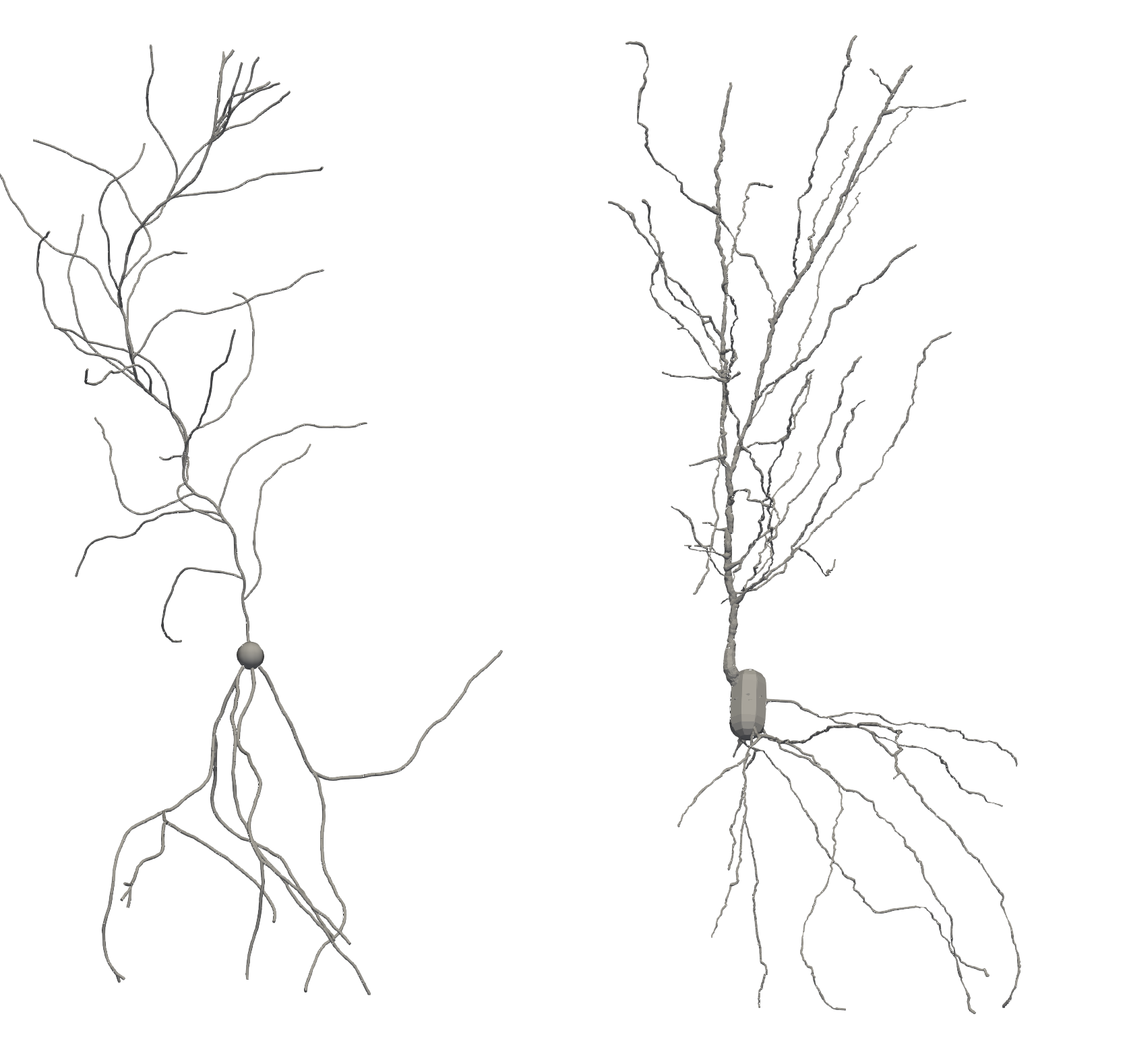}
	}
	\subfloat[neuron-id 23]{
		\includegraphics[width=0.33\textwidth]{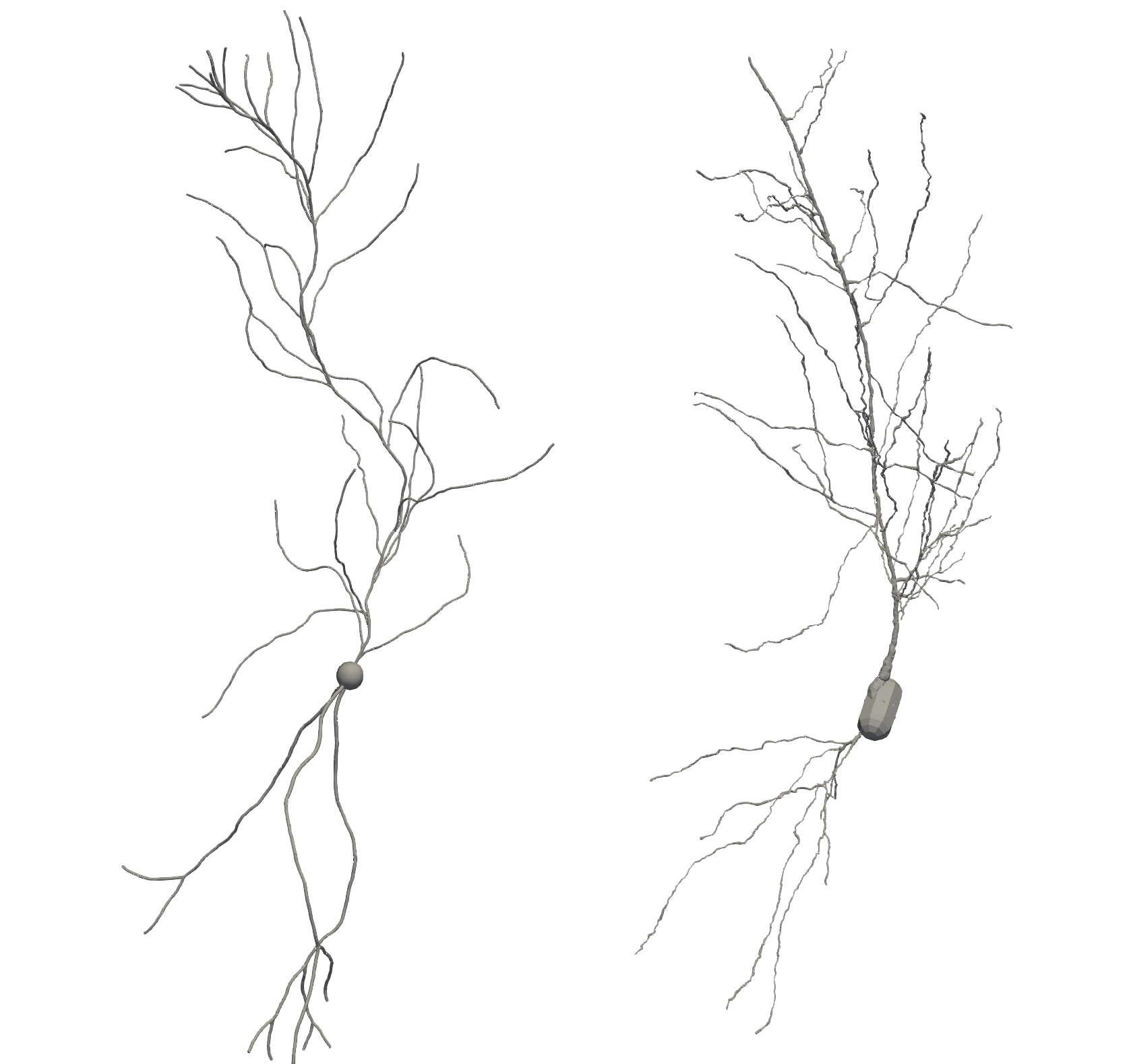}
	} \\
	\subfloat[neuron-id 39]{
		\includegraphics[width=0.33\textwidth]{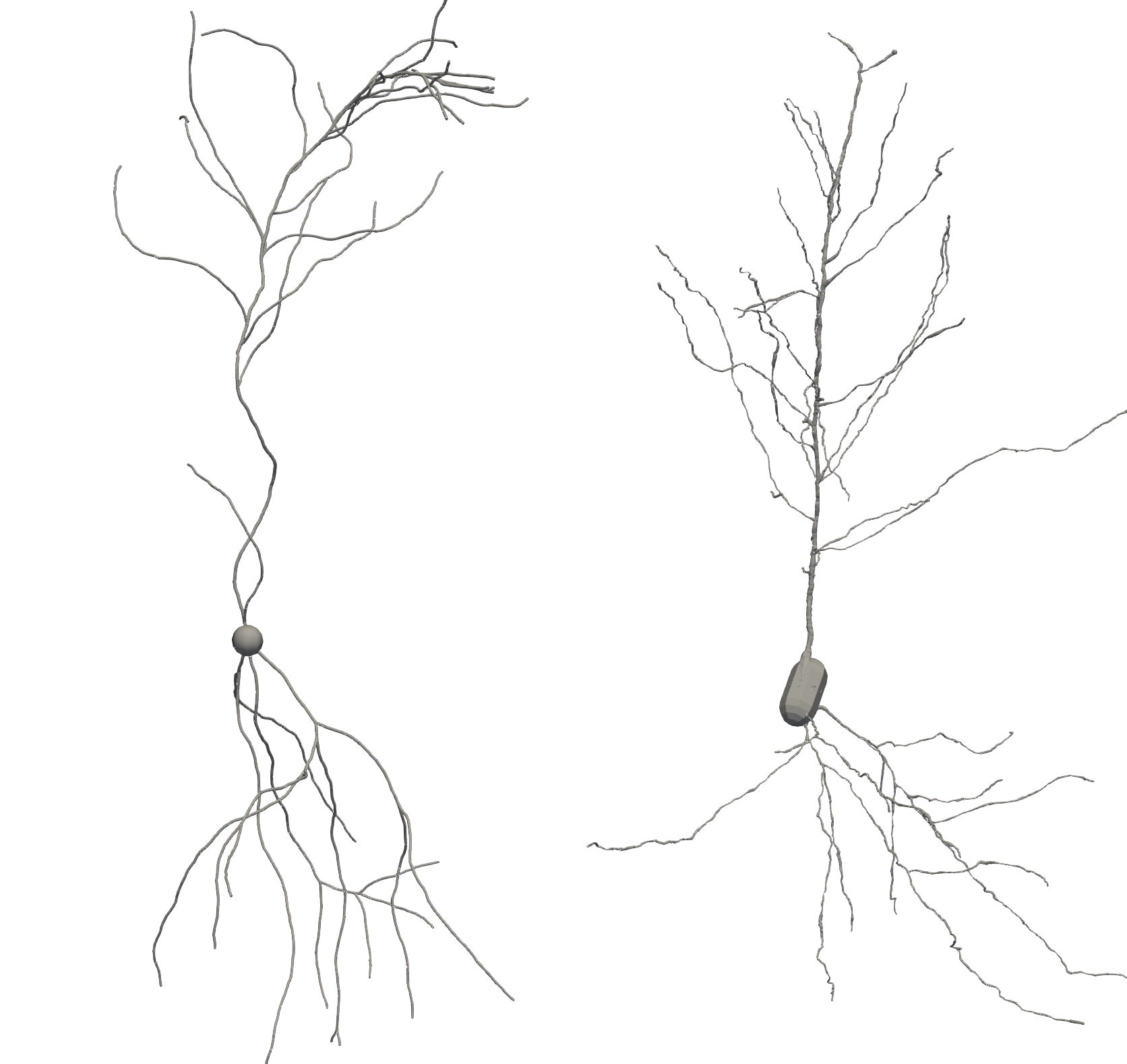}
	}
	\subfloat[neuron-id 43]{
		\includegraphics[width=0.33\textwidth]{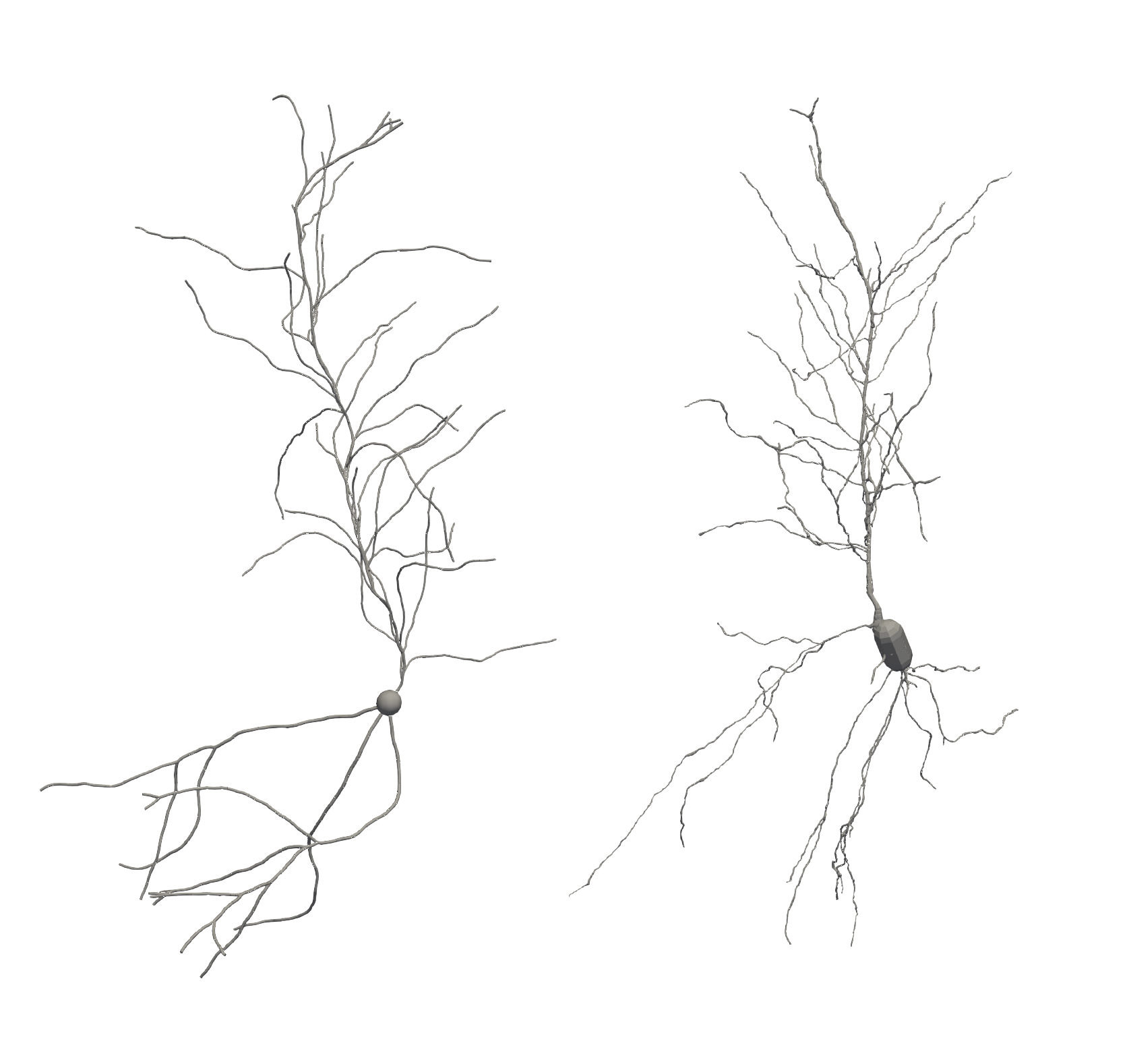}
	}
	\subfloat[neuron-id 46]{
		\includegraphics[width=0.33\textwidth]{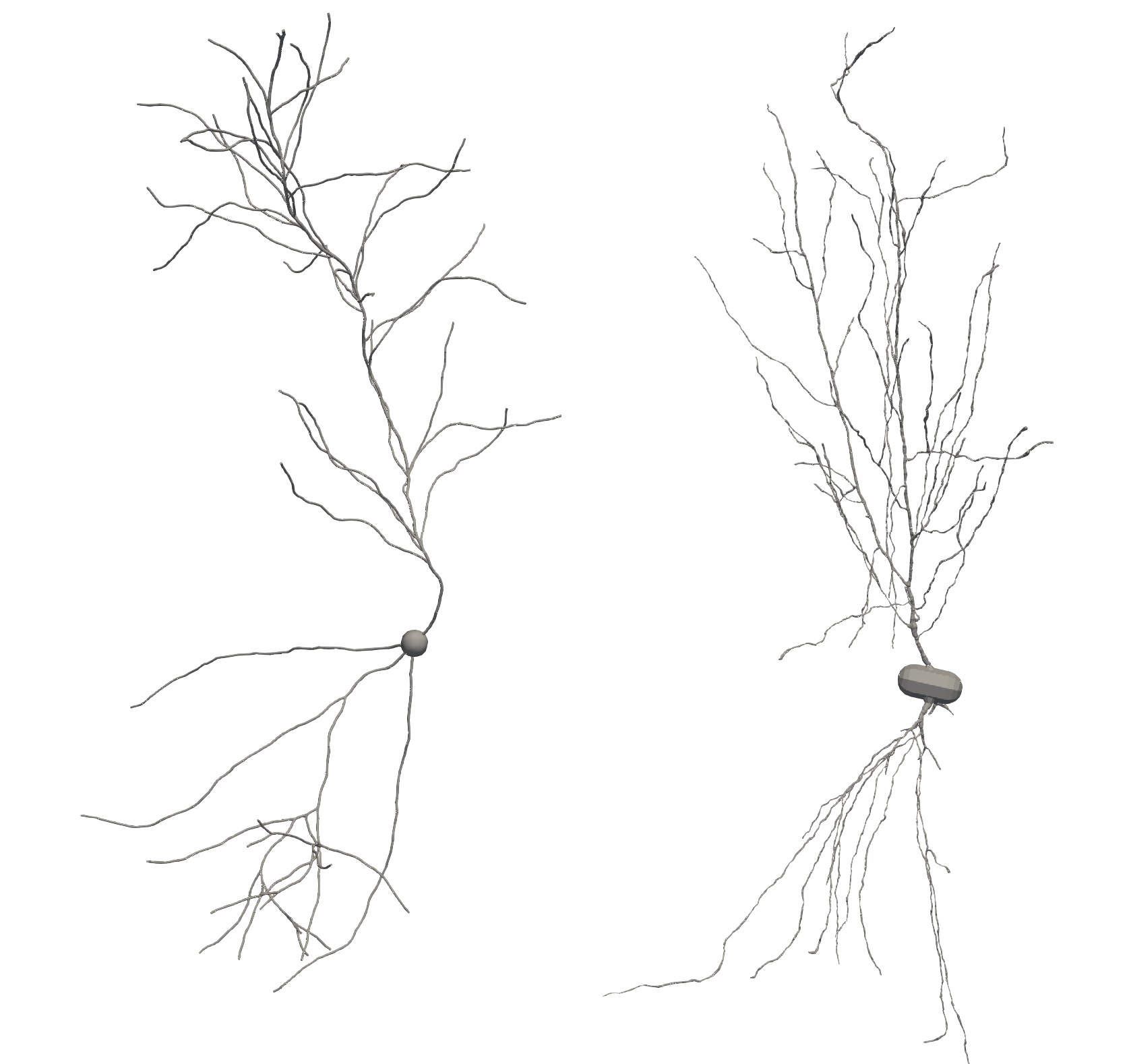}
	} \\
	\subfloat[neuron-id 26]{
		\includegraphics[width=0.33\textwidth]{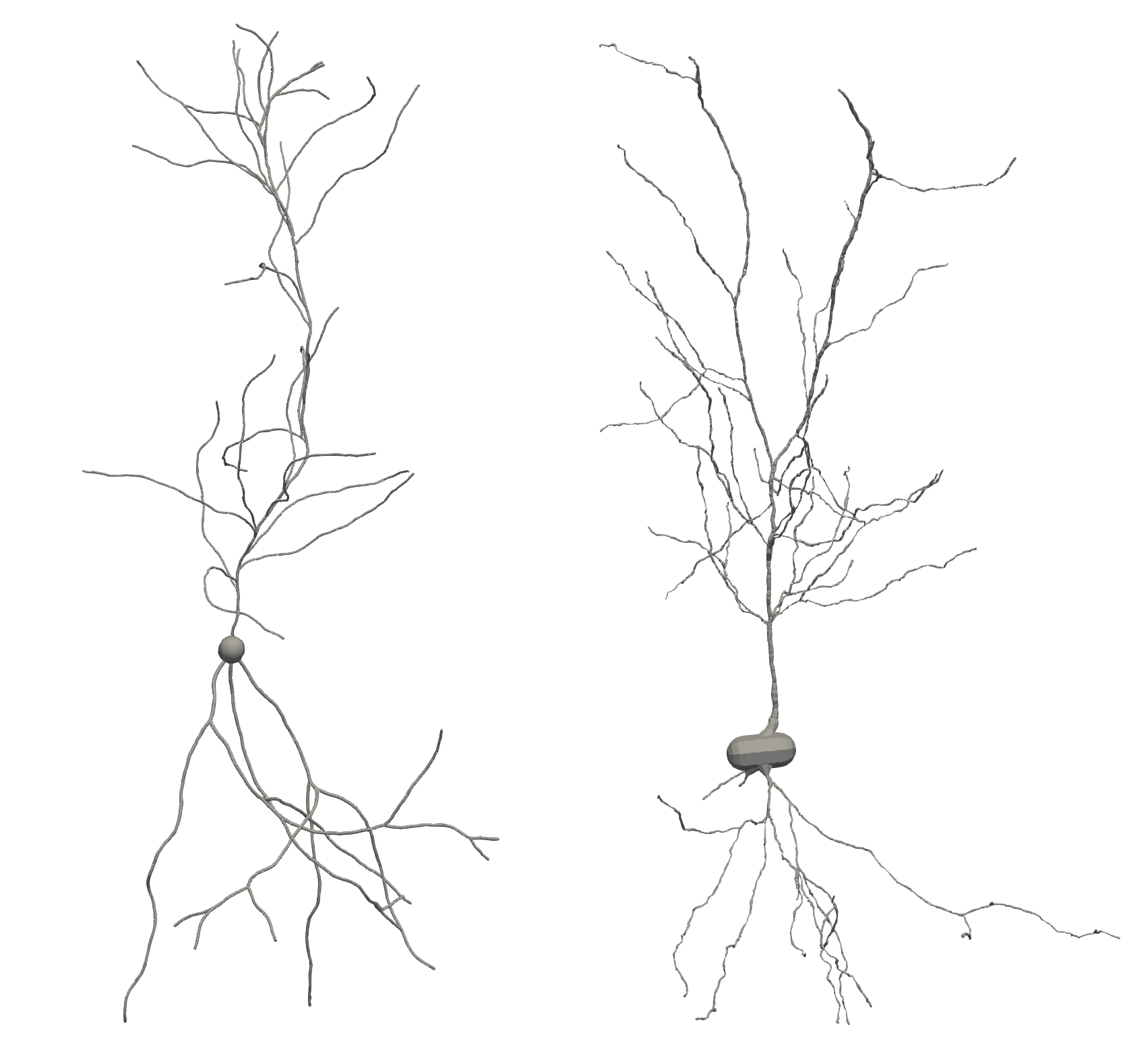}
	}
	\subfloat[neuron-id 29]{
		\includegraphics[width=0.33\textwidth]{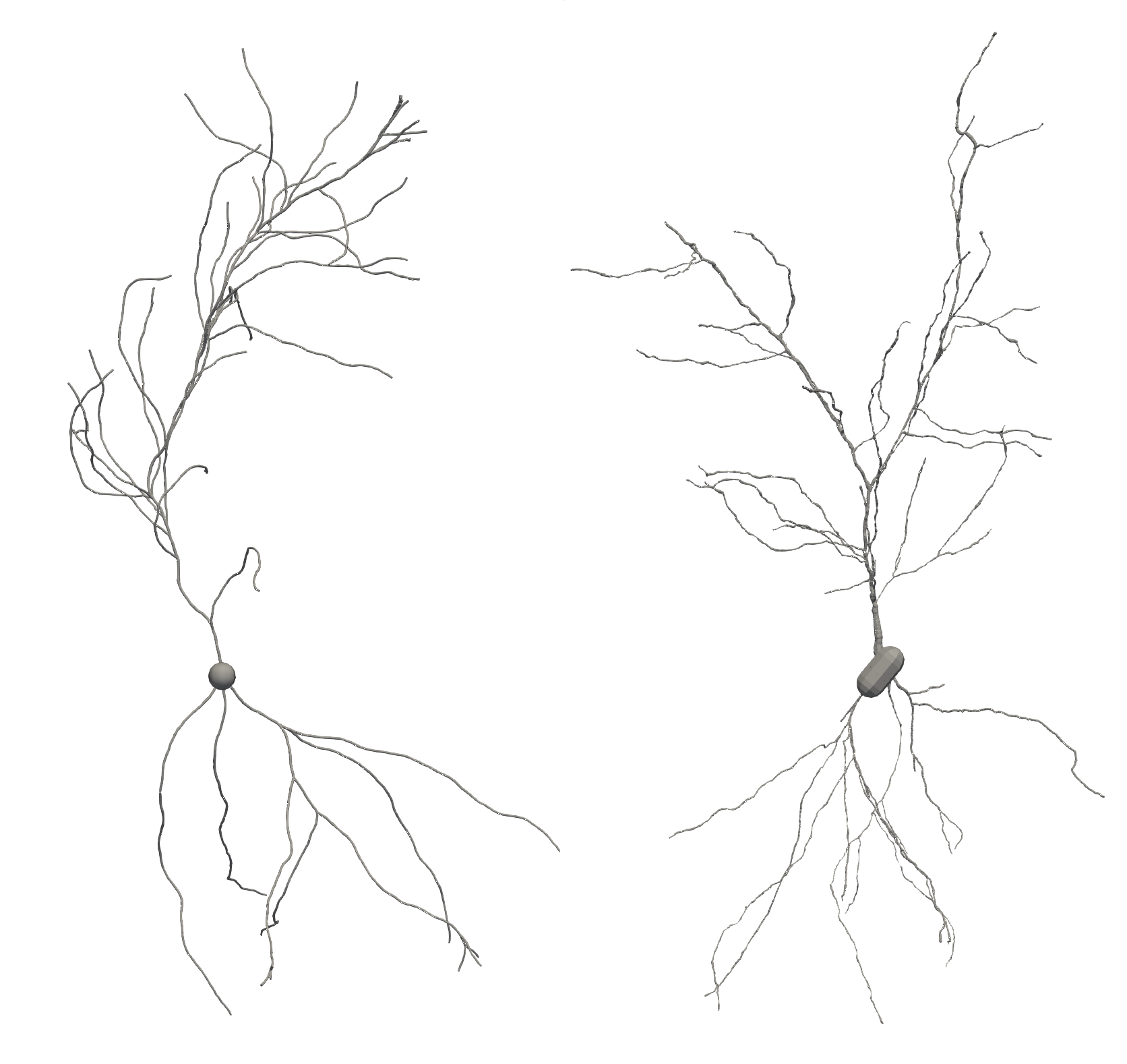}
	}
	\subfloat[neuron-id 37]{
		\includegraphics[width=0.33\textwidth]{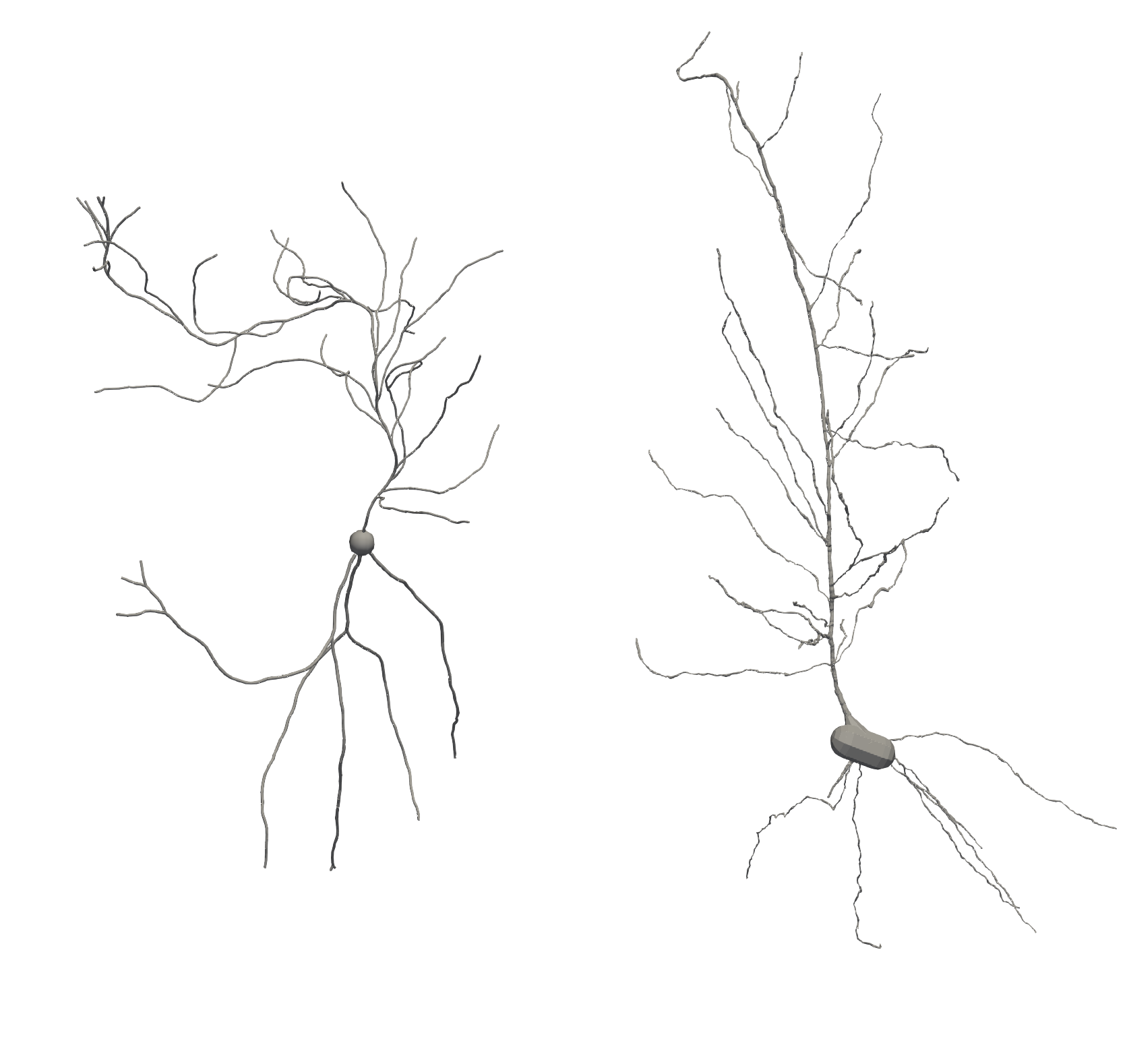}
	}
	\caption{
		Simulated neurons after calibration (left of each panel next) together with
		experimental pyramidal cells~\citep{DeFelipePyramidal2019} (mouse, hippocampus).
		Visualization with ParaView.
	}
	\label{fig:real-vs-sim}
\end{figure}

Figure~\ref{fig:real-vs-sim} compares simulated and experimental pyramidal cells by displaying their full 3D structure. Panels (a)-(f) shows neurons that appear to be similar regarding the characteristic morphology. Closer investigation also reveals differences, in particular, the apical section (top part) of the experimental neurons shows a characteristic, straight main branch. The simulated neurons develop a similar hierarchy, e.g., a main branch with extensions; however, this feature is more emphasized in the experimental neurons and the main branch barely deviates from a straight line.

While (a)-(f) yield good a agreement, some neurons show morphological features that are beyond the QoIs. For instance, in Panel~(g) and~(h), the experimental neurons show a main branch that effectively splits into two. Such features are not captured by the selected set of QoIs; we may not expect that calibration drives the model parameter in a suitable range. Moreover, it is questionable if the simplistic Model~2 can produce such growth. Lastly, Panel~(i) shows an example for which the curvy main branch folds, a feature that can occur in the simulation but is not readily observed in the data. These limitations seem inherent to the simplistic model and can be partially bypassed with different weights for the bias and persistence of the walk.

\subsection{Runtime and computational cost}

We omit detailed benchmarks since they strongly depend on the model, and thus, precise numbers may not be adequate estimates for future research endeavors. Nonetheless, we would like to give a few numbers for ballpark estimates.

The numerical experiments were conducted on two different systems. System~1 is an Intel Xeon E7-8890 CPU with 72 physical cores distributed among four sockets with 1~TB RAM. The system's maximal clock frequency is 3.3 GHz. System~2 is a compute node with two sockets, each hosting an AMD Epyc 7713 CPU with 64 physical cores and 1~TB RAM; hence, a total of 128 cores and 2~TB RAM. The system's maximal clock frequency is 3.7 GHz. On both systems, we run the experiments in Docker containers.

For our calibration, we capped the runs at 50 million simulated neurons. We stop the calibration after the iteration crossing this threshold has finished. For experiments with the Wasserstein distance and synthetic data from Model~2, the calibration took roughly 95 hours on system~1 (using 72 cores) and 62 hours on system~2 (using 96 cores). Taking into consideration the total number of neurons sampled during this time, system~1 and system~2 achieve a throughput of roughly 148 and 225 neurons per second, respectively. Per core, the two systems yield a throughput of 2.06 and 2.35 neurons per second.

We note that the previous measurements include the Wasserstein distance calculation. \citet{Bernton2019} pointed out that computing the Wasserstein distance is expensive compared to evaluating summary statistics and suggested cheaper alternatives such as the Hilbert or swapping distance. However, they considered models that are significantly faster to compute (e.g., a bivariate normal distribution); for the case at hand, we usually simulate datasets of $\nsamples=50$ neurons, which takes approximately 20-25 seconds. Computing the Wasserstein distance numerically as in~\citep{Flamary2021pot, Dutta2021, Bonneel2011} takes significantly less than a second and, thus, the computational time of the Wasserstein distance computation is none of our concerns.

We further remark that, in some cases, we observe substantial imbalances in the computational workload across the MPI ranks (similar to~\cite[Figure~5]{Dutta2021}). For instance, some SMC iterations in the experiment displayed in Figure~\ref{fig:smcabc-convergence-distance-comparison}~(c) ran on only four ranks for extended periods of time. Scalability issues were partially addressed in~\cite[Section~3]{Dutta2021} suggesting using dynamic work sharing in an MPI context. However, the load imbalances seem inherent to the algorithm more than they can be attributed to the implementation.

\section{Discussion}

Before reflecting the individual computational experiments, we wish to emphasize that our experiments collectively demonstrate that ABC combined with morphometrics and statistical distances presents a potent tool for addressing the statistical inverse problem for neuronal growth models. The remainder of the section structurally follows the preceding one.

\subsection{Sensitivities}

We analyzed the influence of parameter variations on the expectation value of the QoIs, utilizing Saltelli's method and computing the Sobol indices. The results showed that almost all QoIs are influenced by a set of parameters and cannot typically be explained by a single parameter in isolation. The discrepancies between the total effect index and the first-order sensitivity index hint at higher-order effects. In sum, this shows that even simple models yield complex emerging behavior.

We recall that sensitivities measure variation and, thus, information. If certain QoIs are sensitive to parameter variations, these QoIs likely contain information that may be leveraged during Bayesian inference to find the posterior distribution of the parameter. On the contrary, QoIs that are insensitive to the model parameters are unlikely to add information that can be leveraged during inference. Such QoIs should be omitted because their statistical fluctuations may pollute the distance metric. Here, our sensitivities verified that the selected morphometrics contain relevant information for inferring the parameters.

\subsection{SMCABC on synthetic data}

We performed various numerical experiments with synthetic data and Model~2, investigating the choice of morphometrics, influences of statistical distances, and the number of samples per parameter. The results are depicted in the Figures~\ref{fig:smcabc-convergence-effect-morphometrics},~\ref{fig:smcabc-convergence-distance-comparison}, and~\ref{fig:smcabc-convergence-effect-datasize}, respectively. We highlight that the experiments clearly demonstrate that ABC with statistical distances and morphometrics can uncover the posterior distributions for simple, resource-driven neuronal growth models.

\paragraph{Effect of the morphometrics.}

To infer parameters, the QoIs must be informative for the considered parameter set. This effect is demonstrated in Figure~\ref{fig:smcabc-convergence-effect-morphometrics} in which we gradually increase the information available for parameter inference by adding one QoI at a time. By leveraging the statistical distances, changing datasets and QoIs becomes trivial endeavor as they automatically consider the distribution without the need of defining appropriate summary statistics for the additional QoIs. This highlights the versatility of the proposed framework and how easy it is to employ in practice. We accentuate that this feature allows the method to easily generalize to more complex neuron models. For a given model, SA helps quantifying which QoIs are sensitive to parameter variations. These QoIs may then be chosen for the inference -- ABC based on statistical distances then automatically accounts for the added information in the comparison step~\eqref{eq:abc_distance_criterion}.

We initialized all experiments in Figure~\ref{fig:smcabc-convergence-effect-morphometrics} with uniform priors. In practical application, Bayesian frameworks support leveraging previous information. As new data becomes available or as new QoIs are computed, the previously determined posterior may serve as a prior to the new inference problem. This technique may be relevant when approaching more intricate models requiring more intensive calculations. We redirect the reader to~\citet{Oden2017} and~\citet{Lima2018} for more details.

\paragraph{Effect of different statistical distances.}

Figure~\ref{fig:smcabc-convergence-distance-comparison} shows the convergence of the posterior marginals for different statistical distances. Despite their different characteristics, the SMCABC sampler~\citep{DelMoral2012, Bernton2019} finds posterior distributions in agreement with the data-generating parameter across most statistical distances. Comparing the Wasserstein distance to the KL and $\gamma$ divergence, we observe that the latter requires fewer SMC iterations to concentrate around the data-generating parameter for the elongation speed and resource consumption. However, the Wasserstein distance better identifies the branching probability. Surprisingly, the sliced Wasserstein performed poorly even though it approximates the Wasserstein distance in high-dimensional settings; it appears as if this approximation failed capture sufficient information in this low-dimensional use-case to infer the parameters. Overall, the Wasserstein distance resulted in the most efficient and reliable calibration algorithm, i.e., it found the best posterior distributions as measured by the density at the data-generating parameters while being reliable in its convergence.

When running with KL and $\gamma$, we encountered issues with the SMCABC sampler. In particular, we aborted both algorithms after exhausting 82\% and 85\% of their simulation budget, respectively, because individual particles got trapped in low probability regions preventing the algorithm from further progressing. Here, we used $\alpha = 0.6$ for all four algorithms, determining how aggressively the SMC sampler moves forward. We suspect that a different choice of $\alpha$ would aid the convergence of these algorithms. A closer investigation of this hypothesis is of interest for the future but outside the scope of this manuscript. Since the purpose of this experiment was to better understand the effects of different distance measures, we fixed all parameters, including $\alpha$, between experiments.

\paragraph{Effect of simulated dataset size.}

The posterior quality of ABC algorithms generally improves with the computing resources that can be allocated to it. The more data available, the better simulations can be compared with data. The more SMC iterations are executed, the better the results. The more simulations are executed per parameter, the more profound the understanding of the model's statistical properties. Moreover, the more simulated and observed data is available, the better the estimates of the statistical distances, see Figure~\ref{fig:wasserstein-gaussian}.

From our experiments in Figure~\ref{fig:smcabc-convergence-effect-datasize}, we conclude that Wasserstein ABC finds the data-generating parameter even when only small sample sizes are used, e.g., $\nsamples = 10$ or $\nsamples = 25$. In the absence of computational limitations, it is clear that larger sample sizes are favorable. We decided to use $\nsamples=50$ samples per parameter for the experiments on experimental data as this seems to be the best compromise between computational runtime and reliability given our computational resources.

\subsection{SMCABC on experimental data}

In both cases, i.e, the apical dendrites of human and mouse pyramidal cells in the hippocampus, Wasserstein ABC finds sharp posterior distributions indicating significant confidence in the parameter estimation, see Figure~\ref{fig:smcabc-convergence-experimental-data-1-apical} and~\ref{fig:smcabc-convergence-experimental-data-2-apical}. The posterior marginals of the mouse data show higher densities and, thus, indicate more confidence in the parameter values. The predictive check shows that the model's prediction and the data agree concerning their mean values but also reveals that the predictions underestimate the width of some QoI distributions, for instance, the standard deviation of the segment lengths. This also applies to the mean segment length of the human pyramidal cells. The datasets show larger variations than the models predictions for these cases.

Figure~\ref{fig:real-vs-sim} presented nine pairs of simulated and experimental neurons to facilitate a more detailed comparison. Some of the pairs, i.e., (a)-(f) underline the similarities between simulation and experiment while others expose shortcoming of the models. We point out the examples~(g) and~(h) for which the main branch of the apical dendrites breaks into two parts. Neither Model~2 nor the selected QoIs reflect this feature, thus, we cannot expect to find it in the calibrated neurons. Model extensions considering random walk weights and branching probabilities depending on the resource parameters bear the potential to reduce this reality gap in future work.

George Box famously described this ubiquitous reality gap between natural phenomena and mathematical models as \textit{all models are wrong, but some are useful}. Whether any model, e.g., Model~1 or~2, is a useful approximation to real neurons depends on the subsequent downstream applications. In other words, different applications require the models to accurately mimic distinct neuronal properties that reflect in a set of carefully chosen, application-dependent QoIs. For instance, simulating the detailed electro-physiology of a single neuron imposes different requirements on synthetic neurons than simulating a cortical region involving thousands of neurons. Researchers implementing a simulation based on synthetic neurons must critically reflect on whether the neuron's QoIs match the problem sufficiently well.

While the models are simplistic and impose a limit on the agreement of data and simulation within the predictive check, the experiment demonstrated that Wasserstein ABC can find the posterior distributions explaining limited features (selected QoIs) of the observed data for two different types of pyramidal cells.

\section{Conclusion}

This investigation explored approximate solutions for the Bayesian inverse problem encountered in the context of neuron growth models. We investigated the ability of morphometrics to extract essential characteristics of neuronal morphology and explored how statistical distances can be effectively used to incorporate this information into SMCABC samplers. To achieve this, we initially focused on simplified mathematical models for illustrative purposes. We embedded these models into a more general abstract concept -- the resource-driven neuron growth model -- drawing upon and summarizing existing literature. Finally, to enhance accessibility and potential adoption of this methodology, we presented an implementation efficiently coupling BioDynaMo and ABCpy.

Our investigations on synthetic data demonstrate the effectiveness of ABC with various statistical distances, e.g., the Wasserstein distance and KL/$\gamma$ divergence, in recovering the data-generating parameters. While Wasserstein yielded superior results within the specific experimental setup, further investigation is necessary to confirm general advantages due to potential bias towards specific parameters, particularly the SMC parameter $\alpha = 0.6$ used in this study. The proposed framework facilitates the seamless integration of additional structural information (morphometrics) into the statistical analysis, demonstrating the flexibility of the approach. Additionally, our findings reveal that the employed SMC sampler achieves satisfactory convergence even with few simulations per parameter. Applications to experimental data confirm successful model calibration through the algorithm. We further find that the simple mathematical models describe data derived from mouse pyramidal cells better than human counterparts. Beyond the analyzed quantities of interest, a comprehensive comparison of complete 3D structures revealed additional similarities and discrepancies between simulation and experiment.

Our framework holds significant promise for advancing research in mechanistic, agent-based neuron growth models and neuroscience. The abstract, resource-driven growth model is a robust foundation for systematically exploring diverse realizations and constructing detailed representations of specific neuron types. This endeavor is supported by the flexibility of the framework pairing morphometrics and statistical distances as it allows for seamless adaptation to new data and relevant quantities of interest. Furthermore, the method paves the way for applying Bayesian model selection and other computational techniques from related fields such as predictive computational science~\citep{Oden2017}. We believe the community can significantly benefit from adopting these ideas; however, as highlighted by \citet{Robert2011}, critical assessment of the approximations inherent in ABC remains crucial when employing the posterior for model selection.

In a broader context, the proposed framework possesses the potential to contribute to the classification of diverse neuronal types by linking the neuron's morphology (captured through morphometrics), its functional role, and mathematical models. Moreover, this approach could support the simulation of large-scale brain structures, enabling the exploration of phenomena such as cortical lamination~\citep{Bauer2021} by calibrating neuron models for different brain regions. Additionally, a proper understanding of the stochastic processes driving neuronal growth may empower the development of heuristic, stochastic, and biologically-inspired algorithms for designing neural network architectures tailored for specific tasks.

While our experiments effectively demonstrate the method's potential, we acknowledge certain limitations in our approach. The utilized models and quantities of interest are intentionally kept simple for illustrative purposes and likely require further refinement for broader applicability. Furthermore, the employed datasets are relatively small, and additional data points would facilitate a more thorough comparison of simulation and experiment. Additionally, the algorithm's runtime scales linearly with the model's runtime. More complex models may require increased computational resources or the exploration of computationally cheaper surrogates (e.g., Gaussian processes~\citep{Rocha2022}) in place of the current models. Lastly, the SMC algorithm occasionally encounters convergence issues (specifically with KL/$\gamma$ divergence) where particles become trapped in low-probability regions.

Addressing the above limitations presents promising directions for future research. In particular, developing more detailed and complex neuron growth models based on the presented abstract concept, as well as their verification with the SMCABC framework, is a crucial step for further improvements. Different models and neuron types beyond pyramidal cells deserve attention and should be the subject of future studies. Extensions considering growth of multiple, interacting neurons or activity dependent growth presents another interesting avenue. With new neuron types and more extensive simulations, additional, more involved morphometrics must be considered when mapping neurons to a low-dimensional vector space. Such simulations may also benefit from hierarchical Bayesian calibration procedures allowing to tackle a sequence of simpler sub-problems~\citep{Oden2017,Lima2018}. Concerning the ABC algorithm, one should address the issue of single particles limiting calibration progress. Intuitively, many failed attempts to update a single particle signal that it is trapped in a low probability region. Quantifying this information and including it in the resampling step seems a promising avenue for operating the algorithm even more smoothly. Lastly, the posterior distribution contains information that can be used to choose between different mathematical models (see OPAL algorithm presented by~\citet{Oden2017}), which would be a substantial step forward. These efforts would collectively support the method's robust application in more complex scenarios.

In conclusion, this work establishes SMCABC based on statistical distances and morphometrics as a potent tool for approximating the solution of the Bayesian inverse problem and, therefore, calibrating neuronal growth models.

\backmatter

\bmhead{Supplementary information}\label{suppl}

The project's code will be shared on GitHub/Zenodo after acceptance of the manuscript.

\bmhead{Data availability}\label{dataavail}

This research did not generate new data. All datasets used in the computational experiments are publicly available and can be accessed through neuromorpho.org.

\bmhead{Acknowledgements}

The work of T.D. has been sponsored by the Wolfgang Gentner Programme of the German Federal Ministry of Education and Research (grant no. 13E18CHA). The work of B.W. was partly funded by the German Research Foundation by grants WO671/11-1. R.B. was supported by the Medical Research Council (grant no. MR/N015037/1) and the Engineering and Physical Sciences Research Council (grant nos. EP/S001433/1 and EP/S001433/2).

\section*{Declarations}

The authors declare no competing interest.

\begin{appendices}

	\section{Sliced-Wasserstein distance}
	\label{sup:s1-calibration-sliced-wasserstein}

	Calibration of Model~2 with the sliced-Wasserstein distance and synthetic data.

	\begin{figure}[H]
		\centering
		\includegraphics[width=\textwidth]{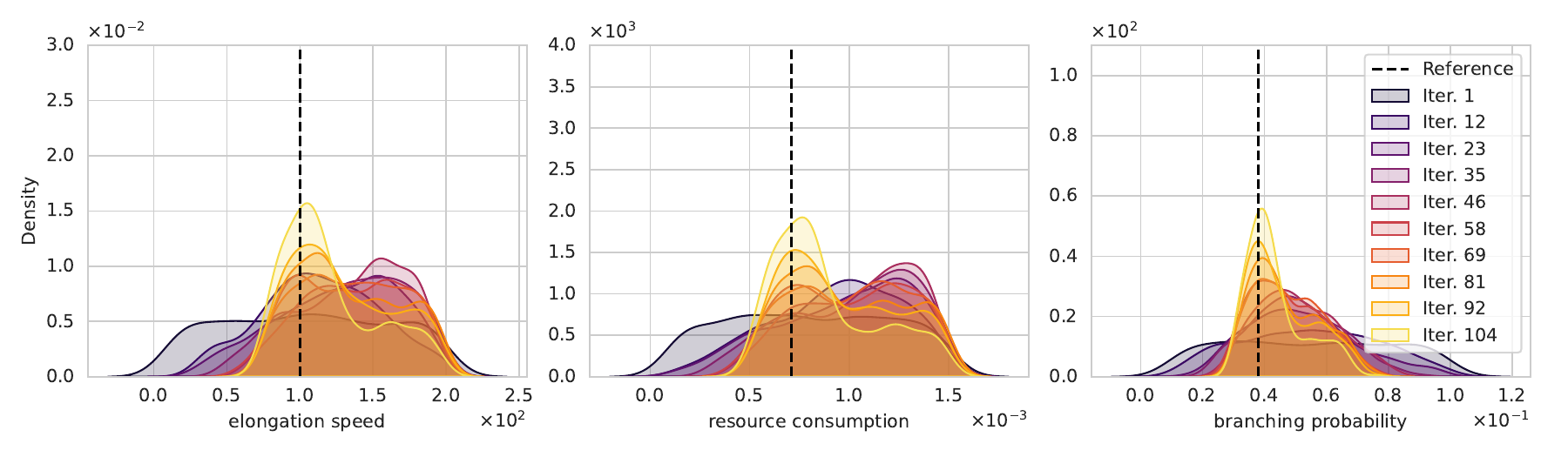}
		\caption{
			Convergence of the ABC algorithms over SMC iterations for the sliced Wasserstein Metric.
		}
		\label{fig:s1-calibration-sliced-wasserstein}
	\end{figure}

	\section{Parameter and Descriptive Statistics}
	\label{ap:parameter}

	\begin{table}[h]
		\caption{Parameter sets for model stochasticity study, see Figure~\ref{fig:model-stochasticity}.}
		\label{tab:parameter-stochasticity-study}
		\begin{tabular}{@{}lc|lc@{}}
			\toprule
			\multicolumn{2}{c|}{\textbf{Model 1}} & \multicolumn{2}{c}{\textbf{Model 2}}                                           \\
			Parameter                             & Value                                & Parameter        & Value                \\ \midrule
			$p_{bra}^\star $                      & $0.006$                              & $p_{bra}^\star $ & $0.038$              \\
			$R^\star$                             & $0.85 \cdot 10^{-3}$                 & $D^\star$        & $0.71 \cdot 10^{-3}$ \\
			$v^\star$                             & $50$                                 & $v^\star$        & $100$                \\ \bottomrule
		\end{tabular}
	\end{table}

	\begin{table}[h]
		\caption{Descriptive statistics of the distributions depicted in Figure~\ref{fig:model-stochasticity}.}
		\label{tab:descriptive_statistics_syndata}
		\begin{tabular}{l|rrrr|rrrr}
			\toprule
			{}   & \multicolumn{4}{c|}{Model 1} & \multicolumn{4}{c}{Model 2}                                                                                                                                                                   \\
			{}   & $\mathcal{M}1$               & $\mathcal{M}2 \ [ \mu m ]$  & $\mathcal{M}3 \ [ \mu m ]$ & $\mathcal{M}4 \ [ \mu m ]$ & $\mathcal{M}1$ & $\mathcal{M}2 \ [ \mu m ]$ & $\mathcal{M}3\ [ \mu m ] $ & $\mathcal{M}4 \ [ \mu m ]$ \\
			\midrule
			mean & 32.44                        & 43.31                       & 34.77                      & 1249.23                    & 38.40          & 60.02                      & 42.80                      & 2279.70                    \\
			std  & 18.14                        & 13.32                       & 7.71                       & 517.77                     & 8.47           & 4.34                       & 2.35                       & 419.44                     \\
			min  & 3.00                         & 21.97                       & 0.01                       & 444.09                     & 11.00          & 42.59                      & 30.73                      & 893.79                     \\
			25\% & 19.00                        & 35.17                       & 29.42                      & 867.12                     & 33.00          & 57.07                      & 41.52                      & 1993.24                    \\
			50\% & 29.00                        & 40.00                       & 33.65                      & 1154.93                    & 39.00          & 59.68                      & 42.79                      & 2256.92                    \\
			75\% & 43.00                        & 47.42                       & 38.90                      & 1525.91                    & 43.00          & 62.43                      & 44.00                      & 2554.55                    \\
			max  & 147.00                       & 148.11                      & 71.07                      & 4840.67                    & 77.00          & 93.85                      & 84.81                      & 4271.21                    \\
			\bottomrule
		\end{tabular}
	\end{table}

	\begin{table}[h]
		\caption{Sensitivity analysis: parameter bounds for $\Omega$ for the different models.}
		\label{tab:parameter-sensitivity-study}
		\begin{tabular}{@{}lcc|lcc@{}}
			\toprule
			\multicolumn{3}{c|}{\textbf{Model 1}} & \multicolumn{3}{c}{\textbf{Model 2}}                                                                               \\
			Parameter                             & min                                  & max                 & Parameter & min                 & max                 \\ \midrule
			$p_{bra} $                            & $0.003$                              & $0.01$              & $p_{bra}$ & $0.003$             & $0.01$              \\
			$R$                                   & $0.4 \cdot 10^{-3}$                  & $1.2 \cdot 10^{-3}$ & $R$       & $0.4 \cdot 10^{-3}$ & $1.2 \cdot 10^{-3}$ \\
			$v$                                   & $30$                                 & $100$               & $v$       & $30$                & $100$               \\ \bottomrule
		\end{tabular}
	\end{table}

        \newpage
	\section{Algorithms.}
	\label{S1_Algorithms}

	\textbf{Simulation logic and agent algorithms.}
	This appendix gives pseudo-code of the simulation logic and the algorithms that govern the agents during the neuronal growth simulation.

	\begin{algorithm}[H]
		\caption{BioDynaMo's simulation protocol.\\
			\textbf{Parameter:} $T$ simulation time, $dt$ time step
		}
		\label{alg:BDM}
		\begin{algorithmic}[0]
			\State $t = 0$
			\While{$t < T$}
			\State Set up Iteration
			\State Propagate Static Information
			\ForAll{Agents $a$}
			\State Update Static Information
			\State Execute Agent Algorithm \Comment{\textbf{Neuron Model:} rules / behaviors} \label{alg:BDM,line:behavior}
			\State Compute Mechanical Forces
			\State Discretize
			\State Propagate Static Information (Agent)
			\EndFor
			\State Tear down Iteration
			\State Update the Environment
			\State Export Visualization and Data
			\State Increment time: $t \gets t + dt$
			\EndWhile
		\end{algorithmic}
	\end{algorithm}

	\begin{algorithm}[H]
		\caption{Model 1.
			The algorithm is embedded in Alg.~\ref{alg:BDM}, line~\ref{alg:BDM,line:behavior}.
			\\
			\textbf{Parameter:} bifurcation probability $p_{bra}$, resource consumption $R$,
			elongation speed $v$, weights $w_{1,2,3}$, external gradient $\nabla \phi$,
			resource threshold $r_{\min}$\\
			\textbf{Agent Attributes:} resource $r$ , if agent is tip $s$ (boolean),
			start point $\vec{x}_s$, end point $\vec{x}_e$, length $l$.
		}
		\label{alg:Model1}
		\begin{algorithmic}[0]
			\If{$s = \text{true}$ \textbf{and} $d > d_{\min}$}
			\State $\vec{d}_1 \sim U(-1,1)^3$
			\State $\vec{d}_2 = (\vec{x}_e - \vec{x}_s) / l$
			\State $\vec{d}_3 = \nabla \phi ((\vec{x}_e + \vec{x}_s)/2)$
			\State $\vec{d}_3 \gets \vec{d}_3 / || \vec{d}_3 ||$
			\State $\vec{y} = \sum_{i=1}^{3} w_i \vec{d}_i$ \Comment{correlated, biased, random walk}
			\State $\vec{x}_e \gets \vec{x}_e + v \vec{y}$ \Comment{elongation}
			\State $r \gets r - R$ \Comment{reduce resource}
			\If{$z \sim U(0,1) < p_{bra}$}
			\State Trigger Bifurcation
			\EndIf
			\EndIf
		\end{algorithmic}
	\end{algorithm}

	\begin{algorithm}[H]
		\caption{Model 2.
			The algorithm is embedded in Alg.~\ref{alg:BDM}, line~\ref{alg:BDM,line:behavior}.
			\\
			\textbf{Parameter:} branching probability $p_{ba}$, resource consumption $R$,
			elongation speed $v$, weights $w_{1,2,3}$, external gradient $\nabla \phi$,
			resource threshold $r_{\min}$, start resource $r_0$ \\
			\textbf{Agent Attributes:} resource $r$ , if agent is tip $s$ (boolean),
			start point $\vec{x}_s$, end point $\vec{x}_e$, length $l$.
		}
		\label{alg:Model2}
		\begin{algorithmic}[0]
			\If{$d > d_{\min}$}
			\State $\vec{d}_1 \sim U(-1,1)^3$
			\State $\vec{d}_2 = (\vec{x}_e - \vec{x}_s) / l$
			\State $\vec{d}_3 = \nabla \phi ((\vec{x}_e + \vec{x}_s)/2)$
			\State $\vec{d}_3 \gets \vec{d}_3 / || \vec{d}_3 ||$
			\State $\vec{y} = \sum_{i=1}^{3} w_i \vec{d}_i$ \Comment{correlated, biased, random walk}
			\State $\vec{x}_e \gets \vec{x}_e + v \vec{y}$ \Comment{elongation}
			\State $r \gets r - R$ \Comment{reduce resource}
			\If{$z \sim U(0,1) < p_{bra}$ \textbf{and} $s = \text{true}$}
			\State Trigger Branch
			\State Set resource of new branch to $r_0$
			\EndIf
			\EndIf
		\end{algorithmic}
	\end{algorithm}

\end{appendices}

\bibliography{bibliography}%

\end{document}